\documentclass[superscriptaddress,nofootinbib,onecolumn,12pt]{revtex4-2}

\usepackage[paperwidth=210mm,paperheight=297mm,centering,hmargin=2cm,vmargin=2.5cm]{geometry}

\usepackage{amssymb}
\usepackage{amsmath}
\usepackage{graphicx}
\usepackage{mathdots}

\usepackage{verbatim}

\newcommand{\be}{\begin{equation}}
\newcommand{\ee}{\end{equation}}
\newcommand{\bea}{\begin{eqnarray}}
\newcommand{\eea}{\end{eqnarray}}

\newcommand{\ii}{\mathrm{i}}

\begin{document}
 \title{Landau-Zener transition rates of superconducting qubits
 and absorption spectrum in quantum dots}

\author{Jorge G. Russo}\email{jorge.russo@icrea.cat}
\affiliation{Instituci\'o Catalana de Recerca i Estudis Avan\c{c}ats (ICREA), \\ Pg.~Lluis Companys, 23, 08010 Barcelona, Spain}\affiliation{Departament de F\' \i sica Cu\' antica i Astrof\'\i sica and Institut de Ci\`encies del Cosmos, \\ Universitat de Barcelona, Mart\'i Franqu\`es, 1, 08028 Barcelona, Spain }
\author{Miguel Tierz}\email{mtierz@ucm.es}
\affiliation{Departamento de An\'alisis Matem\'atico y Matem\'atica Aplicada, Universidad Complutense de Madrid, 28040 Madrid, Spain}
 
\begin{abstract}
New exact formulas are derived for systems involving Landau-Zener transition rates and for absorption spectra in quantum dots. These rectify previous inaccurate approximations utilized in experimental studies. The exact formulas give an explicit expression for the maxima and minima of the transition rate at any oscillating period and reveal a number of striking physical consequences, such as the suppression of oscillations for half-integer values of the detuning parameter and that the periodic dependence on the detuning parameter changes at special values of the driving field amplitude. The fluorescence spectra of quantum dots exhibit similar properties.
\end{abstract}

\maketitle



\section{Introduction}

The production of a coherent superposition between quantum states in nanoelectronic
circuits is one of the cornerstones for the development of quantum technology \cite{krantz2019quantum}. A distinguished example of this is the Landau-Zener-Stückelberg-Majorana (LZSM) interferometry \cite{shevchenko2010landau,ivakhnenko2023nonadiabatic}, which formerly was also simply known as Landau-Zener interferometry.
The LZSM theory, was first developed around 1932 in the context of the study of spin dynamics and slow atomic collisions,
demonstrated that transitions are possible between two
approaching levels as a control parameter is swept across
the point of minimum energy splitting. For a detailed account of all the different aspects and nuances of the first four seminal contributions, see \cite{di2005majorana,ivakhnenko2023nonadiabatic,kofman2023majorana,nikitin1999nonadiabatic}. 

In LZSM interferometry, a quantum two-level system is driven strongly across an
avoided energy-level crossing producing first a quantum superposition between the two states  of the system. The temporal evolution of the two states occurs with different dynamical phases and,
after a second passage through the anticrossing, coherent interference between these two states occurs.
A periodically driven field can also induce coherent Rabi oscillations 
for a different regime of parameters.
Rabi oscillations  occur when the driving frequency $\nu=\omega/2\pi $ is of the same order
as the energy-level separation $\Delta E/\hbar$. The rate equation-based description,
where transitions occur through the Landau-Zener effect, arises at $\hbar\nu\ll \Delta E$ and strong driving amplitude $A\sim\Delta E$ (for discussions, see \cite{ivakhnenko2023nonadiabatic,berns2006coherent}).

There is an analogy between LZSM interferometry and Mach-Zehnder interferometry since the beamsplitters can be realized by Landau-Zener (LZ) transitions at a level avoided crossing and, over one oscillation period of the driving field, the qubit is swept through the avoided crossing twice. This point of view is developed in, for example, \cite{oliver2005mach,ivakhnenko2023nonadiabatic} and it is applicable to settings discussed below, such as the double-passage setting (which is related to optical Mach-Zehnder \cite{ivakhnenko2023nonadiabatic}).

LZSM interferometry has been studied in a number of platforms. In the first part of this paper, we will be focusing on works where the platform is based on superconducting Josephson junctions \cite{berns2006coherent,rudner2008quantum} (see also \cite{li2013motional}). These works are characterized by the presence of noise but Landau-Zener transitions in externally driven systems have been considered, for example, also under dissipation \cite{gefen1987zener1,shimshoni1991onset} or in the context of other mesoscopic systems \cite{gefen1987zener2}.

We shall thoroughly reanalyze the transition rate in \cite{berns2006coherent,rudner2008quantum}, providing an exact analytical expression for it as well as an improved asymptotic characterization of the rate, which will include a more accurate portrayal of the oscillations with full analytical control.

The mathematical formula describing the LZSM transition rate also appears 
in the study of the absorption spectrum of quantum dots, with a reinterpretation of the parameters. In the second part we shall examine  settings involving various external fields.
An example  consists in a system describing the dipole coupling  of the quantum dot to a laser field.
Concretely, we consider  InAs/GaAs quantum dots (nanoscale islands of InAs embedded in a GaAs matrix) studied in   \cite{metcalfe2010resolved}
and study the absorption spectrum for two different configurations.

We shall also study the two-level system with modulated laser light discussed in \cite{blind1980resonance}, 
which  appeared prior to the discovery of quantum dots, but it is nevertheless governed by related physics. 
Finally, we will examine the power spectrum of quantum dots in presence of a bichromatic electromagnetic field, investigated in \cite{kryuchkyan2017resonance},
which is expressed as an infinite series of Mollow triplets \cite{mollow1969power}. 
Here the corresponding infinite sums will be computed analytically.

\section{Analysis of the transition rate}

Consider the two-level system (TLS) studied in \cite{berns2006coherent}
of a periodic driving on a qubit, with Hamiltonian
\begin{equation}
    {\cal H} = -\frac{1 }{2} \left(\begin{matrix}
h(t) & \Delta \\
\Delta  & -h(t) 
\end{matrix}\right)\ ,
\end{equation} 
where $h(t)=\mathcal{E} +\delta
{\cal E} (t)+A\cos \omega t$ is the energy detuning (or bias)
from an avoided crossing, modulated by the driving
field. The term $\delta
\mathcal{E}(t)$ describes classical noise and  $\Delta$ is the energy gap between the two levels.
The LZ transition rate was computed in perturbation theory
in \cite{berns2006coherent}, with the result
\begin{equation}
\label{uno}
    W=\frac{\Delta^2}{2}\sum_{n=-\infty}^\infty \frac{\Gamma_2 J_n^2(x)}{(\mathcal{E}-\omega n)^2+\Gamma_2^2}\ ,\qquad x\equiv \frac{A}{\omega}\ ,
\end{equation}
where $J_{n}$ is the Bessel function of the first kind. 
Note the symmetry property $W(-\mathcal{E})=W(\mathcal{E})$. The parameter $\mathcal{E}$ is the energy bias and $\Gamma_2$ is the decoherence rate \cite{berns2006coherent,ivakhnenko2023nonadiabatic}\footnote{If  the relaxation time $T_1$ is taken into account, the upper level occupation probability can be calculated as in \cite{ivakhnenko2023nonadiabatic} (see section $3.6$ and Appendix B.4.2). The resulting expression (Eq. B.62 in \cite{ivakhnenko2023nonadiabatic}), the stationary solution of the rate equation, cannot be evaluated using our analytic formulas.}.

The sum in equation \eqref{uno} is usually evaluated numerically. However,
attempts have been made to provide analytic
approximations, as analytic formulas can reveal properties that are 
difficult to identify using numerical methods with generic parameters.
In particular, in \cite{berns2006coherent}, and later on in \cite{otxoa2019quantum,ivakhnenko2023nonadiabatic}, this sum was 
computed by assuming that the main contributions come from large $n$ terms, one can replace 
\begin{equation}
\label{tres}
    J_n(x)\approx a {\rm Ai}[a(n-x)]\ ,\quad a=(2/x)^{1/3}\ .
\end{equation}
This approximation works well for a few oscillations but then the two functions $J_n(x)$ and $ a {\rm Ai}[a(n-x)]$ get out of phase.

  The assumption that the main contributions come from large $n$ terms requires that $\mathcal{E}\gg \omega$, so that 
 the sum is dominated by  terms with $n\approx n_0\equiv [\mathcal{E}/\omega]$.

In  \cite{berns2006coherent}, the remaining sum was approximated by the following  heuristic formula 
\begin{equation}
\label{Wval}
W_{\rm app}= \frac{\pi a^2\Delta^2}{2\omega} {\rm Im}\big[\cot(\pi \mu^*)\big]\, {\rm Ai^2}\big[\frac{a}{\omega}(\mathcal{E}-A)\big]\ ,
 \qquad \mu\equiv \frac{1}{\omega}(\mathcal{E}+i\Gamma_2)\ .
\end{equation}
A numerical comparison shows that
this formula  approximates the sum \eqref{uno} provided
\begin{itemize}
    \item $\frac{\mathcal{E}}{\omega}\approx {\rm integer}\gg 1$.

    \item $\Gamma_2\ll 1$.
\end{itemize}
A comparison between the original expression \eqref{uno} and the approximation \eqref{Wval} is shown in fig. 1.
We see big deviations between $W$ and $W_{\rm app}$ because in the figures 
$\mathcal{E}/\omega$ is not an integer.
Since physically there is no reason
for $\mathcal{E}/\omega$ to be an integer, generically  $W_{\rm app}$  significantly deviates from the actual rate $W$.
This evident deviation does not affect fig. 4 in   \cite{berns2006coherent}, as the fits with experimental data in that figure have been carried out by means of a numerical evaluation of the sum \eqref{uno}.

At large $A$, one can use the asymptotic formula for the Airy function,
giving
\begin{equation}
    W_{\rm app}\approx \frac{\Delta^2}{2x \omega} {\rm Im}\big[\cot(\pi \mu^*)\big]\, (1-\frac{\varepsilon}{x})^{-\frac12}
    \cos^2 \left[ \frac{2\sqrt{2}}{3}x (1-\frac{\varepsilon}{x}-\frac{\pi}{4})\right]\ .
    \label{mmap}
\end{equation}
As will become clear in the following, this formula gives an incorrect asymptotic frequency.

The formula \eqref{Wval} reproduces some qualitative features of the actual rate $W$: in particular, $W_{\rm app}$ is very small for $A<\mathcal{E}$ (with $\mathcal{E}\gg \omega$), so transitions occur only for $A> O(\mathcal{E})$.
The frequencies of oscillations are similar. But it should be noted that minima and maxima are very different and also oscillations get out of phase after a few periods.
Importantly,  $W_{\rm app}$ has an infinite number of zeroes as a function of $x$, but the exact $W$ never vanishes for any finite $\Gamma_2$ and  generic values of ${\cal E}$.  
The limit $\Gamma_2\to 0$ will be discussed below.

\medskip

We will now show that one can actually compute the sum \eqref{uno} {\it exactly} and in closed form, using a summation formula derived in \cite{newberger1982new} (reviewed in the appendix). 
Consider in general $\mu=\mu_1+i\mu_2$. We first split
\begin{equation}
    \sum_n \frac{J_n(x)^2}{(n-\mu_1)^2+\mu_2^2}=\frac{1}{2i\mu_2 }\left(\sum_n \frac{J_n^2(x)}{ n - \mu}-\sum_n \frac{J_n^2(x)}{ n -\mu^*}
     \right)\,  .
\end{equation}
Using the formula \eqref{neuber} of the appendix we thus find
\begin{equation}
\label{Wexacta0}
 \sum_n \frac{J_n(x)^2}{(n-\mu_1)^2+\mu_2^2}=-\frac{1}{2i\mu_2 } \left(\frac{\pi}{\sin(\pi\mu) }\, J_\mu (x) J_{-\mu} (x)-\frac{\pi}{\sin(\pi\mu^*) }\, J_{\mu}^* (x) J_{-\mu}^* (x)\right)\ .
\end{equation}
or
\begin{equation}
\label{Wexacta}
\boxed{
 \sum_n \frac{J_n(x)^2}{(n-\mu_1)^2+\mu_2^2}=-\frac{1}{\mu_2}{\rm Im}  \left(\frac{\pi}{\sin(\pi\mu) }\, J_\mu (x) J_{-\mu} (x)\right)
}
\end{equation}
where we used that $J_{\mu^*}(x)=J_\mu^*(x)$ for real $x$.
Therefore
\begin{equation}
\label{Wexacta2}
W=-\frac{ \Delta^2}{2\omega}{\rm Im}  \left(\frac{\pi}{\sin(\pi\mu) }\, J_\mu (x) J_{-\mu} (x)\right)\ .
\end{equation}
A numerical check shows that indeed the formula \eqref{Wexacta2}
exactly reproduces the expression \eqref{uno} given by the infinite sum.

\medskip

\begin{figure}[h!]
 \centering
 \begin{tabular}{cc}
\includegraphics[width=0.4\textwidth]{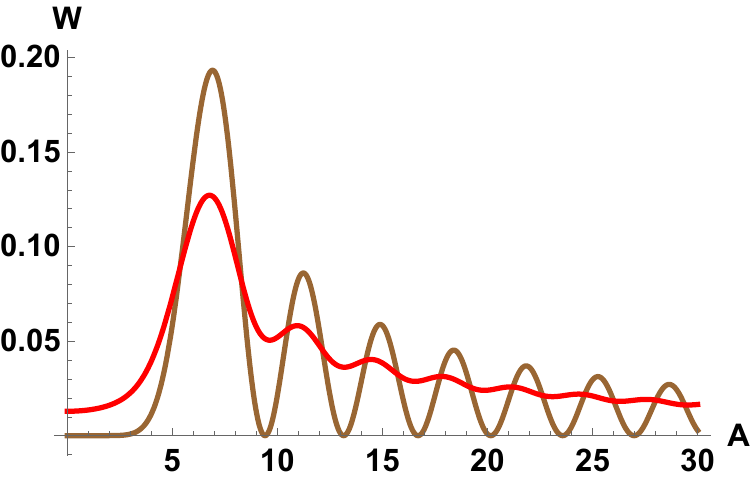}
 &
 \qquad \includegraphics[width=0.4\textwidth]{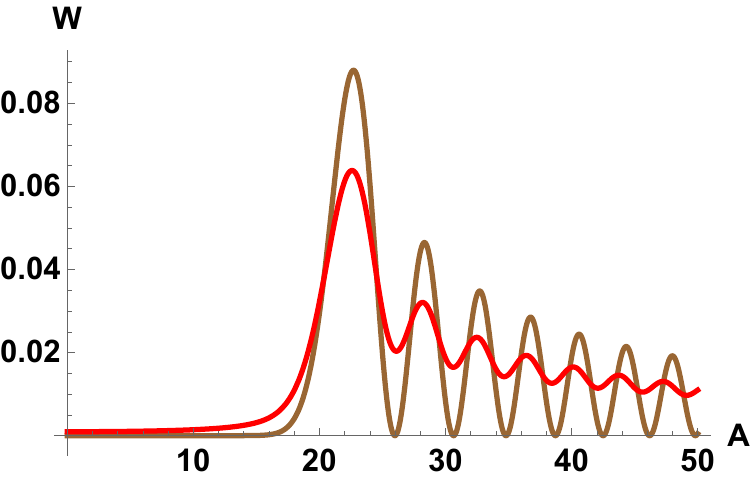}
 \\ (a)&(b)
 \end{tabular}
 \caption
 {(a) $W$ (in units of $\Delta^2/\omega$) as  a function of $A$ (in units of $\omega)$. The approximation \eqref{Wval} (brown) vs. the exact formula \eqref{Wexacta2} (red). Here  $\mathcal{E}=5.5\, \omega $, $\Gamma_2=5 \omega/(2\pi)$.
  (b) $\mathcal{E}=20.5\, \omega $, same value for $\Gamma_2$.
 }
 \label{figab}
 \end{figure}

Modulo an overall factor $\Delta^2/\omega$, $W$ depends on three parameters $(\varepsilon, \gamma, x)$, with 
$$
\varepsilon\equiv\frac{\mathcal{E}}{\omega}\ ,\ \ \gamma \equiv\frac{\Gamma_2}{\omega}\ ,\quad x\equiv \frac{A}{\omega}\ .
$$
We can set $\omega=1$ and plot the rate $W$ as a function of $A$ for different values of ${\cal E}$ and $\Gamma_2$, as in fig. 1.
As noted above, the most notable difference is that  the approximated transition rate \eqref{Wval}  vanishes for specific values of $x$.
This disagrees with the fact that the actual rate \eqref{uno}, \eqref{Wexacta2} does not vanish 
anywhere for generic values of $\Gamma_2 $ and ${\cal E}$.

The transition rate \eqref{uno} exhibits an infinite set of resonances created by the driving force, which appear at integer values of the detuning $\varepsilon$.
All  resonances are  encapsulated in the exact formula \eqref{Wexacta2}.
Indeed,  at  the special values ${\rm Re}[\mu]=n$, $n\in \mathbb{Z}$, one has
\begin{equation}
   - {\rm Im}  \left(\frac{\pi}{\sin(\pi\mu) }\, J_\mu (x) J_{-\mu} (x)\right)=(-1)^n \frac{\pi}{\sinh(\pi\gamma )} \ {\rm Re}\left[ J_{n+i\gamma}(x) J_{-n-i\gamma} (x)\right]\ .
\end{equation}
This shows that for integer $\varepsilon$ the transition rate undergoes the expected enhancement at small $\gamma$, 
\begin{equation}
   - {\rm Im}  \left(\frac{\pi}{\sin(\pi\mu) }\, J_\mu (x) J_{-\mu} (x)\right)\approx  \frac{1}{\gamma } \ J_{n}^2 (x) +O(\gamma)\ ,
\label{gtt}
\end{equation}
a result that is also evident from the original formula \eqref{uno}, where the term $\mathcal{E}=\omega n$ dominates.
This is in contrast with the behavior for generic (non-integer) values of $\varepsilon$, where the
transition rate is proportional to $\gamma $.

Equation \eqref{gtt} shows that the height of the resonance peak 
is proportional to $J_{n}^2 (x)$ and  is therefore a function of the amplitude $A=x\omega$. For special values of the amplitude where
the $J_n(x)$ has a zero, the peak at the detuning value $\varepsilon={\rm Re}[\mu]=n$ disappears, but other peaks with $\varepsilon= n'$, with $n'\neq n$ remain.
Since the Bessel function possesses an infinite number of zeroes, it is apparent that there are infinite special values of amplitude where the peak at any given detuning $\varepsilon =n$ is suppressed.

Equation \eqref{gtt} also clarifies that the transition rate may  have zeroes under two conditions: $\gamma\to 0$ and integer $\varepsilon$.
In these cases, the zeroes of the (rescaled) transition rate $\gamma W$ coincide with the zeroes of the Bessel function $J_n(x)$. 

\medskip

A simple formula for $W$ can also be given for  large $x$, where one has the well known Bessel asymptotics \cite{watson1922treatise},
\begin{equation}
\label{asyJ}
    J_\mu(x)\approx \frac{\sqrt{2}}{\sqrt{\pi x}}\cos(x-\mu\frac{\pi}{2}-\frac{\pi}{4})\ ,\qquad \ x\gg {\rm Max}\{ 1,|\mu|^2 \}\ .
\end{equation}

\begin{figure}[h!]
 \centering
 \begin{tabular}{cc}
\includegraphics[width=0.4\textwidth]{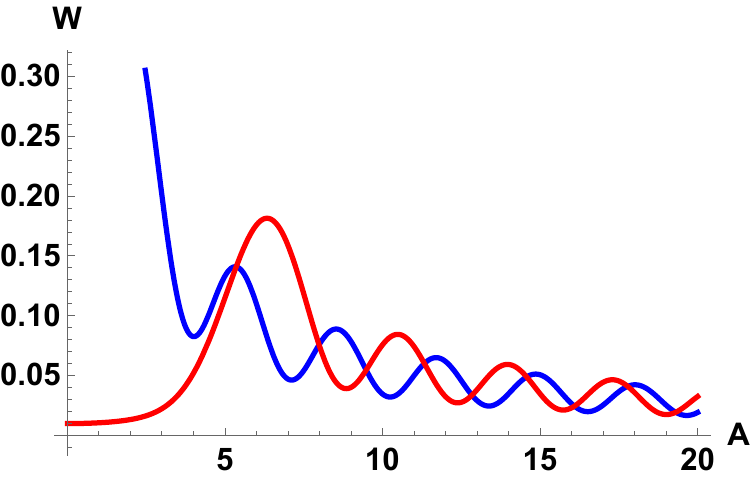}
 &
 \qquad \includegraphics[width=0.4\textwidth]{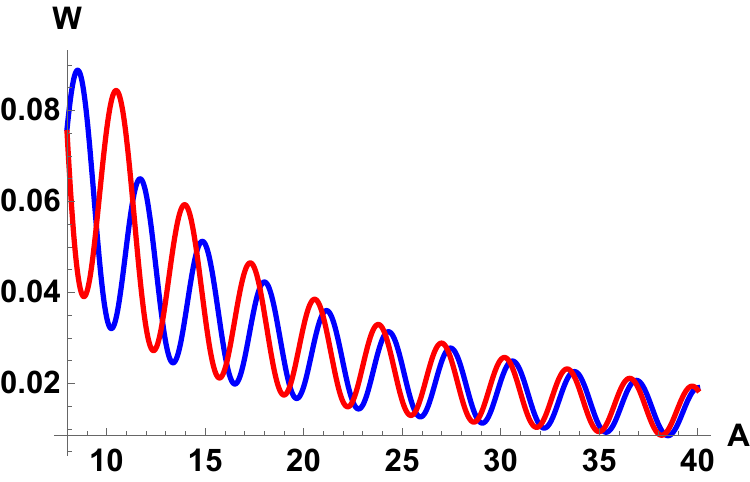}
 \\ (a)&(b)
 \end{tabular}
 \caption
 {At large amplitudes, the behavior of the transition rate is governed  by the simple asymptotic
 behavior of the Bessel functions (red and blue colors correspond to $W$ and $W_{\rm asym}$).  Here  $\varepsilon =5$, $\gamma=0.5 $, and $W$ is given in units of $\Delta^2/\omega$. 
 (a) At small $x$, one can see a gap due to the well known  turning point of the Bessel functions around $x\sim|\mu|$ ({\it i.e.} $A\sim\mathcal{E}$). (b) When $x\gg |\mu|^2$, the asymptotic formula approaches the exact result.
 }
 \label{fig2ab}
 \end{figure}
 
Thus, for large $x$ 
\begin{equation}
\label{Wlargex}
-{\rm Im}  \left(\frac{\pi}{\sin(\pi\mu) }\, J_\mu (x) J_{-\mu} (x)\right) \approx -\frac{1}{x} {\rm Im}\left(\frac{\cos(\pi\mu)+\sin(2x) }{\sin(\pi\mu)}\right)\ .
\end{equation}
 Computing the imaginary part, we find
 
\begin{equation}
-{\rm Im}  \left(\frac{\pi}{\sin(\pi\mu) }\, J_\mu (x) J_{-\mu} (x)\right) \approx \frac{2}{x}\, \frac{\sinh(\pi \gamma )}{\cosh(2\pi\gamma )-\cos(2\pi\varepsilon)} \left(\cosh(\pi \gamma)+ \cos(\pi\varepsilon)\sin(2x)\right)\ .
\label{JJasym}
\end{equation}

\noindent Thus
\begin{equation}
\label{Wlargexx}
W_{\rm asym}= \frac{\Delta^2}{x\omega}  \, \frac{\sinh(\pi \gamma )}{\cosh(2\pi\gamma )-\cos(2\pi\varepsilon)} \left(\cosh(\pi \gamma)+ \cos(\pi\varepsilon)\sin(2x)\right)\ , 
\end{equation}
which is clearly different from \eqref{Wval}, \eqref{mmap}.
This formula explicitly shows the correct asymptotic frequency of oscillations in the $x=A/\omega$ variable, given by the factor $\sin(2x)$,
which therefore differs from the frequency in the old approximation \eqref{mmap}.
It also provides an explicit expression for the amplitude of oscillations, which  can be read from the above formula.
A comparison between the exact $W$ and $W_{\rm asym}$ is shown in fig. \ref{fig2ab}.

The minimum and maximum values of the transition rate at any oscillating period can also be read from \eqref{Wlargexx}:
\begin{equation}
    W_{\rm asym}\bigg|_{\rm max}= \frac{\Delta^2}{2x\omega}  \, \frac{\sinh(\pi \gamma )}{\cosh(\pi\gamma )-\cos(\pi\varepsilon)} \ ,\qquad  W_{\rm asym}\bigg|_{\rm min}= \frac{\Delta^2}{2x\omega}  \, \frac{\sinh(\pi \gamma )}{\cosh(\pi\gamma )+\cos(\pi\varepsilon)} \ ,
    \label{vvnn}
\end{equation}
where we assumed $\cos(\pi\varepsilon)>0$. As $\cos(\pi \varepsilon )$ flips sign,  all minima become  maxima and viceversa.

\begin{figure}[h!]
 \centering
\includegraphics[width=0.5\textwidth]{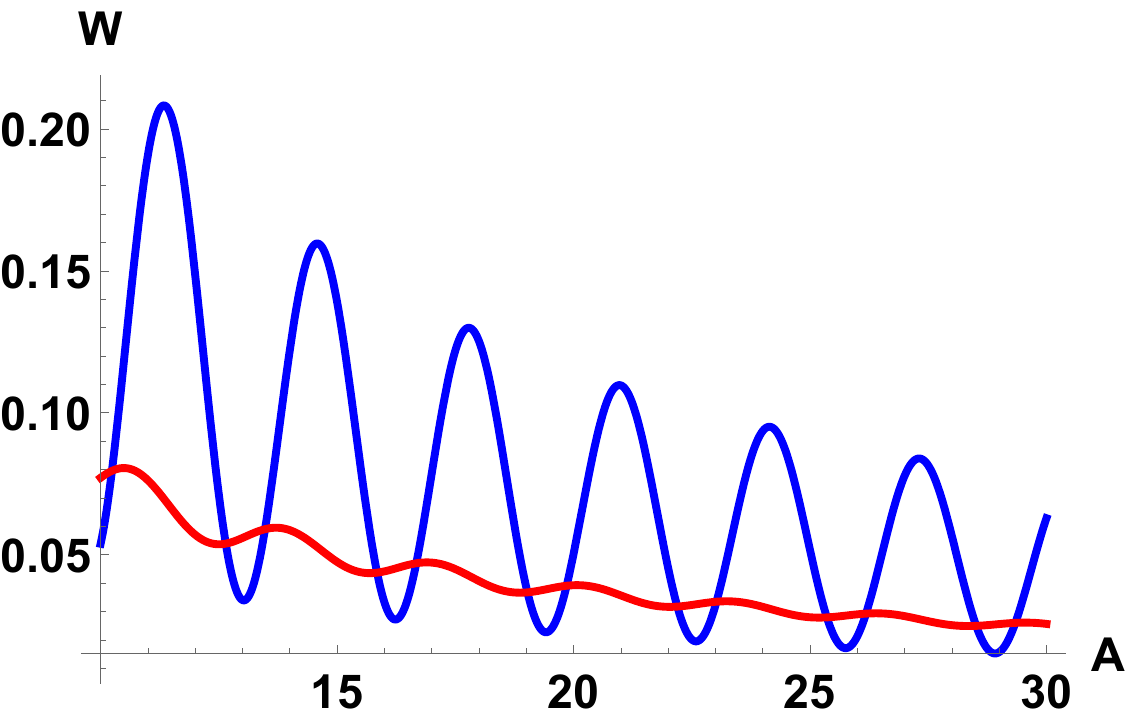}
 \caption
 {The exact $W$ given in \eqref{Wexacta2} (in units of $\Delta^2/\omega$) as  a function of $A$ (in units of $\omega$). The oscillations are strongly suppressed
 at half-integer values of the detuning parameter. Here $\gamma=0.3$, $\varepsilon=5/2$ (red) and $\varepsilon =1.2\times 5/2$ (blue).}
 \label{figwww}
 \end{figure}

The general formula \eqref{Wexacta} can be applied in many experimental setups involving harmonic periodic driving. Some examples will be discussed below. The asymptotic formula  \eqref{Wlargexx} reveals some remarkable features.
To leading order in the asymptotic expansion in $1/x$, the oscillations in the $x$ variable are multiplied by $\cos(\pi\varepsilon)$, so they
vanish for
 the special values of $\varepsilon$,
$$
\varepsilon=n+\frac12,\ \ n\in \mathbb{Z}\ ,
$$
giving
\begin{equation}
  \frac{2}{x}\, \frac{\sinh(\pi \gamma )}{\cosh(2\pi\gamma )-\cos(2\pi\varepsilon)} \left(\cosh(\pi \gamma)+ \cos(\pi\varepsilon)\sin(2x)\right)\ \longrightarrow\ 
    \frac{1}{x}\, \tanh(\pi \gamma)\ .
\end{equation}
For such special values of the detuning parameter ${\cal E}=\varepsilon \omega$ there is a significant suppression in the amplitude of oscillations in $x$, which becomes of $O(1/x^2)$ (the residual oscillating $O(1/x^2)$ contribution originates from corrections
to the asymptotic formula \eqref{asyJ}). This is shown in fig.   \ref{figwww}.
This important property is only revealed after carrying out  the exact summation over $n$, which may explain why it was overlooked in previous analyses.

Another noteworthy property of the formula \eqref{Wlargexx}
regards the $\gamma $ dependence.
For small $\gamma $ and generic values of $\varepsilon$, 
the transition rate is $O(\gamma )$ due to the factor $\sinh(\pi\gamma)$. 
As discussed above, the expected resonances appear for integer $\varepsilon$,  a property that is also manifest  from the asymptotic formula:
\begin{equation}
 W_{\rm asym}\ \longrightarrow\  \frac{\Delta^2}{2x\omega}  \,
    \frac{1}{\sinh(\pi \gamma)}\left(\cosh(\pi\gamma )\pm \sin(2x)\right)\approx \frac{1}{\gamma}\, \frac{\Delta^2}{2\pi x\omega}
    \left(1\pm \sin(2x)\right),
    \nonumber
\end{equation}
where the signs $\pm $ correspond to even and odd $\varepsilon$, respectively. Thus the transition rate becomes large,  $O(1/\gamma )$.
An exception arises for the special values of the amplitude where $\sin(2x)=\mp 1$, resulting in the suppression of the resonance peaks.
These specific values of the amplitude $A=\omega x$ correspond to the asymptotic values of the zeroes of $J_n(x)$, as previously discussed.

Note the marked change of behavior at special values of $x$. For generic $x$, the asymptotic transition rate is a periodic function of $\varepsilon $ with period $\varepsilon=\varepsilon + 2n$.
However, at the special values
$$
x=\frac{n\pi}{2}\ ,\quad n\in \mathbb{Z}\ ,
$$
one has $\sin(2x)=0$ and the period suddenly becomes $\varepsilon=\varepsilon + n$.

\medskip

It would be extremely interesting to experimentally test all these distinctive features.

\medskip

Finally, it is also worth looking at the behavior of $W$ at small amplitudes. Using \cite{hall1965ion}\footnote{The formula follows from an identity in \cite{watson1922treatise}.}
\begin{equation}
\frac{\pi}{\sin(\pi\mu) }\, J_\mu (x) J_{-\mu} (x)=\frac{1}{\mu}\left(1+\sum_{m=1}^\infty \frac{(2m)!}{2^{2m} (m!)^2}
\frac{x^{2m}}{(\mu^2-1^2)(\mu^2-2^2)\cdots (\mu^2-m^2)}\right)\ ,
\end{equation}
we find
\begin{equation}
    W(x= 0) = \frac{\Delta^2 \gamma }{2\omega |\mu|^2} \ .
\end{equation}
This formula is exact and encompasses the entire summation in \eqref{uno}.
This non-vanishing value is clearly visible in fig. 1a (where $W(x=0)\cong 0.013$). This contrasts  the
exponentially small value inaccurately predicted by the old and commonly used approximation \eqref{Wval}.

A number of studies extend the study of transitions to multiple energy levels
and use $W$ as building block as the transition rate between a given pair of energy levels \cite{wen2009landau,liul2023interferometry,liul2023rate}. This introduces of course, multiple energy bias parameters, one for each pair, and our evaluation applies directly to each $W$.

\subsection{Upper-level occupation probability and Stückelberg phase}

The formula \eqref{Wlargexx} can also be written as
\begin{equation}
\label{Wlargexxx}
W_{\rm asym}=  \frac{\Delta^2}{x\omega} b \left(a -  \sin^2(x-\frac{\pi}{2}\varepsilon-\frac{\pi}{4})-  \sin^2(x+\frac{\pi}{2}\varepsilon-\frac{\pi}{4})\right),
\end{equation}
with
\begin{equation}
   b \equiv\frac{\sinh(\pi \gamma )}{\cosh(2\pi\gamma )-\cos(2\pi\varepsilon)} \ ,\ \ a\equiv 2\cosh^2(\frac12 \pi \gamma)\ .
\end{equation}

It should be noted that the oscillation frequency in the $x=A/\omega$ variable is equal to 2. This is in sharp contrast with the predicted asymptotic frequency of the approximation \eqref{Wval},
where, for large $A$, it has the form \eqref{mmap} and
 predicts an incorrect frequency $\frac{4\sqrt{2}}{3}\neq 2$.
As a consequence, at large $x$, the approximation \eqref{Wval} gets completely out of phase.

The above asymptotic behavior \eqref{Wlargexxx} can be compared with known results for the upper-level occupation probability, obtained within the adiabatic-impulse model, in the double-passage regime and in the fast passage limit
$\Delta^2/(A\omega)\ll 1$ \cite{ivakhnenko2023nonadiabatic}. The upper-level occupation probability is the square of the absolute value of the probability amplitude and its time-average is a sum of Lorentzian-shape $k$-photon resonances of higher analytical complexity than the transition rate, see Eq. (55) in \cite{shevchenko2010landau}, but in the above limit, it is proportionally related to the transition rate, giving (see Appendix B$4.4$ in \cite{ivakhnenko2023nonadiabatic})
\begin{equation}
    P_{+}^{\rm double}\approx \frac{2\pi\Delta^2}{\omega A}\ \sin^2\left(x-\frac{\pi}{2}\varepsilon -\frac{\pi}{4}\right)\ .
\end{equation}
Thus we see that the asymptotic expression \eqref{Wlargexxx} corresponding to the exact transition rate
reproduces the expected frequency and also the expected $\varepsilon$
dependence. That is, it reproduces the Stückelberg phase.

\subsection{Fourier transforms of the transition rate}

We now compute the Fourier transforms of $W$ in both variables $\mathcal{E}$ and $A$. 
The characteristic function of a Lorentzian is an
exponential. That is:%
\[
\int_{-\infty }^{\infty }\frac{\lambda e^{ixt}}{\pi \left( \lambda
^{2}+(x-x_{o})^{2}\right) }\ dx=\exp \left( x_{0}it-\lambda \left\vert
t\right\vert \right) .
\]%
We want to Fourier transform with regards to $%
\mathcal{E} $ both in itself and en route to the 2d Fourier, discussed by other means in \cite{rudner2008quantum}. So, we consider
\[
\sum_{n\in\mathbb{Z}
}\frac{J_{n}(x)^{2}\Gamma _{2}}{\left( \mathcal{E} -n\omega \right)
^{2}+\Gamma _{2}^{2}}
\]%
and instead of using \eqref{Wexacta2} and then transform, we carry out the
term by term transform (the dominated convergence theorem
allows us to exchange integration and summation), then%
\begin{equation}
\widehat{W}(k_{\mathcal{E} },x)=
\frac{\Delta^2}{2}\sum_{n\in 
\mathbb{Z}
}\int_{-\infty }^{\infty }\frac{J_{n}(x)^{2}\Gamma _{2}\ e^{-i\mathcal{E}
k_{\mathcal{E} }}}{\left(
\mathcal{E} -n\omega \right) ^{2}+\Gamma _{2}^{2}}\ d\mathcal{E}=\frac{\pi \Delta^2}{2} \sum_{n\in 
\mathbb{Z}
}J_{n}(x)^{2}\exp \left(- n\omega ik_{\mathcal{E} }-\Gamma _{2}\left\vert
k_{\mathcal{E} }\right\vert \right) .  \label{1}
\end{equation}%
Notice that $k_{\mathcal{E}}$ has dimension of time.
The summation can be carried out. Recall Graf's addition
theorem \cite{watson1922treatise}:%
\[
\sum_{l\in 
\mathbb{Z}
}t^{l}J_{l}(x)J_{l+m}(y)=\left( \frac{y-x/t}{y-xt}\right) ^{n/2}J_{m}\left( 
\sqrt{x^{2}+y^{2}-xy\frac{t^{2}+1}{t}}\right) .
\]%
It simplifies since $m=0$ in our case: 
\begin{equation}
\widehat{W}(k_{\mathcal{E} },x)=\frac{\pi \Delta^2}{2} \exp \left( -\Gamma _{2}\left\vert
k_{\mathcal{E} }\right\vert \right) J_{0}\left( \sqrt{2x^{2}(1-\cos \left(
\omega k_{\mathcal{E} }\right) )}\right) \ ,  \label{2}
\end{equation}%
{\it i.e.}
\begin{equation}
\boxed{
\widehat{W}(k_{\mathcal{E} },x)=\frac{\pi \Delta^2}{2} \exp \left( -\Gamma _{2}\left\vert
k_{\mathcal{E} }\right\vert \right) J_{0}\left( 2x \big|\sin\left(
\omega k_{\mathcal{E} }/2 \right)\big|\right) }  \label{2q}
\end{equation}%

The formula shows that $\widehat{W}(k_{\mathcal{E} },x)$
has an infinite number of zeroes in the real $x$ axes, given by the zeroes of the Bessel function.  $\widehat{W}(k_{\mathcal{E} },x)$ exhibits an oscillatory behavior, 
with an amplitude  modulated by the exponential factor 
$\exp \left( -\Gamma _{2}\left\vert
k_{\mathcal{E} }\right\vert \right) $.
In particular, the formula \eqref{2q} predicts the following oscillatory
behavior at small $x$:
\begin{equation}
\label{foux}
\widehat{W}(k_{\mathcal{E} },x)\approx
\frac{\pi \Delta^2}{2} \exp \left( -\Gamma _{2}\left\vert
k_{\mathcal{E} }\right\vert \right) \left(1 -x^2 \sin^2\left(
\frac{\omega k_{\mathcal{E} }}{2} \right)\right)  \ ,\qquad x\ll 1\ .
\end{equation}%

\medskip

We can now look for the
Fourier transform with regards to the $x$ variable (the amplitude) using either (\ref{1}) and/or (\ref{2}). The
route (\ref{2}) seems simpler since we only have to transform $%
J_{0}\left( \alpha x\right) $ where $\alpha :=2\big|\sin\left(
\omega k_{\mathcal{E} }/2 \right)\big|$ and we know that \cite{beca1980orthogonal}:%
\[\int_{-\infty }^{\infty }J_{n}(x)e^{-ik_{x}x}dx=\frac{%
2(-i)^{n}T_{n}(k_{x})}{\sqrt{1-k_{x}^{2}}}\text{ for }\left\vert
k_{x}\right\vert <1,
\]%
where $T_{n}$ is the first degree Chebyshev polynomials (which will not
appear because $n=0$ above). Then%
\begin{equation}
\widetilde{\widehat{W}}(k_{\mathcal{E} },k_{x})= \pi \Delta^2 \frac{\exp \left( -\Gamma
_{2}\left\vert k_{\mathcal{E} }\right\vert \right) }{\alpha \sqrt{%
1-k_{x}^{2}/\alpha ^{2}}}= \pi \Delta^2\frac{\exp \left( -\Gamma _{2}\left\vert
k_{\mathcal{E} }\right\vert \right) }{\sqrt{2\left(1-\cos ( \omega
k_{\mathcal{E} })\right) -k_{x}^{2}}} \ .\label{3}
\end{equation}
Recalling that $x=A/\omega$, in terms of $A$ this reads
\begin{equation}
  \widetilde{\widehat{W}}(k_{\mathcal{E} },k_{A})  =\pi \Delta^2\frac{\exp \left( -\Gamma
_{2}\left\vert k_{\mathcal{E} }\right\vert \right) }{\sqrt{\frac{4}{\omega ^{2}}\sin ^{2}\left( 
\frac{\omega k_{\mathcal{E} }}{2}\right) -k_{A}^{2}}} \ .
\end{equation}
This formula agrees with the double Fourier transform in $\mathcal{E}$ and $A$ given in (15) in \cite{rudner2008quantum}.
However, here we find a new closed formula \eqref{2q} for the single Fourier transform in $\mathcal{E}$
(and, below, also for the Fourier transform in $A$, $\widehat{W}(\mathcal{E},k_{x})$).

\subsubsection{Fourier transform in $A$}


The other partial (1D) Fourier transform, w.r.t. the variable $x$ can also be evaluated.
The function $J_{\mu }(x)J_{-\mu }(x)$ is even in $x$, therefore
(Gradshteyn, 6.672, 2. Page 719, 8th edition)%
\begin{equation}
\widehat{W}(\mathcal{E},k_{x})=\frac{\pi }{\sin \pi \mu }\int\limits_{-\infty }^{\infty }J_{\mu
}(x)J_{-\mu }(x)e^{ixk_{x}}dx =\frac{2\pi }{\sin \pi \mu }\int\limits_{0}^{%
\infty }J_{\mu }(x)J_{-\mu }(x)\cos \left( k_{x}x\right) dx \ .
\end{equation}
Computing the integral we find
\begin{equation}
\widehat{W}(\mathcal{E},k_{x})= \begin{cases} \frac{\pi }{\sin \pi \mu }P_{\mu -\frac{1}{2}}\left( \frac{k_{x}^{2}}{2}%
-1\right)\ \quad &{\rm for\ } 0\leq k_x\leq 2 \\
0 \quad  &{\rm for\ }\ k_x<0\ {\rm or} \ k_x>2
\end{cases}
\end{equation}
where $P_{\mu -\frac{1}{2}}$ is the Legendre function.
Thus the Fourier transform gives a compact support function:
the Fourier transformed transition vanishes identically outside the interval $0< k_x< 2$.

A version of the Payley-Wiener theorem guarantees precisely that: the Fourier transform of a function which is the real restriction of a complex entire function is a function of compact support and vice versa. The Bessel
function $J_{\mu }(z)$ is indeed an entire function. 


\section{Fluorescence spectra of quantum dots}

An examination of literature and reviews on Floquet physics, specifically quantum systems under external AC (harmonic) field modulation, which includes a large number of photo-assisted processes, will reveal a plethora of appearances of expressions of the type on the left-hand side of Eq. \eqref{general}.
In previous studies, this expression was  invariably left in summation form and customarily analyzed only numerically \cite{silveri2017quantum}. 
A prominent example is the power spectrum of fluorescence in quantum dots.

We recall that the transition rate in a modulated system is given by the Fourier transform of the stationary two-time correlation function of the  Hamiltonian
expressed in the interaction picture. Therefore, it is given by the power spectral density of the probe operator evaluated in the unperturbed system. 
Absorptive transitions in the rotating wave approximation 
(RWA)  can be written as \cite{silveri2017quantum}
\begin{equation}\label{eq:SpecQubit}
S(\omega)=\frac{g_{\rm P}^2}{4}\int_{-\infty}^{\infty}e^{\ii \omega t}\langle \hat{\sigma}_-(t)\hat{\sigma}_+(0)\rangle d{t} ,
\end{equation} 
%
%
where $g_P$ represents the strength of the probe and $\hat{\sigma}_\pm$ are qubit raising and lowering operators. 
For harmonic modulations, this leads to the following multi-photon sideband-spectrum:
\begin{equation}\label{eq:rwaspec}
S(\omega) = \sum_{m=-\infty}^{\infty}\frac{g_{\rm P}^2}{2}\frac{\Gamma_2\ J_m^2(x)}{(\delta+m\Omega)^2+\Gamma_2^2}\ ,
\end{equation}
where $\delta\equiv \omega_0-\omega$ is the detuning between the static qubit and the probe. Thus one finds the same expression arising in the previous section in the context of superconducting qubits.  
Indeed, this formula controls the power spectrum in a broad class of two-level systems with harmonic driving \cite{silveri2017quantum,silveri2015stuckelberg}.


\medskip 

Let us now consider the setting of  \cite{metcalfe2010resolved}, studying InAs/GaAs quantum dots.   The formula \eqref{eq:rwaspec}
reappears in the form of resolved
sideband emission, due to surface acoustic waves (SAW) \cite{metcalfe2010resolved}. Indeed, some of the results in \cite{metcalfe2010resolved} are formally equivalent to the previous LZSM results. We will first present the equivalent model and subsequently 
 discuss the effect of adding an interaction term to the Hamiltonian.
This leads to a distinct application of the formula \eqref{general}.

The quantum dot (QD) is modelled as a two-level system (TLS) in \cite{metcalfe2010resolved}, with electric dipole operator $\hat{d}=d\sigma_x$, and dynamics
governed by the Hamiltonian 
\begin{equation}
H=\frac{\hbar}{2}\left(\omega_0+\chi\omega_s \sin(\omega_s t)\right)\sigma_z
\label{eqn: TLS}
\end{equation}
and relaxation terms that cause the off-diagonal elements
of the density matrix $\rho$ to decay at a rate~$\gamma$.
The $\sigma_i$ are Pauli  matrices and the modulation index $\chi $ is a
dimensionless parameter that expresses the amplitude in units of $\omega_s$.

The fluorescence is proportional to the expectation value $\langle \hat{d(t)}\rangle ={\rm Tr}\left(\rho(t)\,\hat{d}\right)$.
In the limit $\omega_s\gg 2\gamma$, where  $2\gamma$ denotes the line width,
one finds the following
power spectrum of the fluorescence  \cite{metcalfe2010resolved}
\begin{equation}
P[\omega]=\sum_{n=-\infty}^{\infty}\frac{J_n^2(\chi)}{\gamma^2+(\omega-\omega_0+n\,\omega_s)^2}\ .
\label{eqn: approxPspect}
\end{equation} 
Applying the summation formula  \eqref{Wexacta}, we find
\begin{equation}
P[\omega]=-\frac{1}{\gamma \omega_s}
{\rm Im}\left( \frac{\pi}{\sin(\pi\nu) }\, J_\nu (\chi) J_{-\nu} (\chi)\right) \ ,\qquad \nu =\frac{1}{\omega_s}(\omega-\omega_0+i\gamma )\ .
\label{aaff}
\end{equation} 
For large $\chi $, we can express this result in terms of 
trigonometric functions, as in \eqref{JJasym}, and read the frequency thereof. We get
\begin{equation}
\label{Wlargepp}
P[\omega]_{\rm asym}=  \frac{2}{\chi\gamma \omega_s} b \left(\cosh(\frac{\pi \gamma}{\omega_s})+ \cos(\frac{\pi}{\omega_s}(\omega-\omega_0) )\sin(2\chi)\right)\ ,
\end{equation}
where now
\begin{equation}
   b \equiv\frac{\sinh(\pi \gamma/\omega_s )}{\cosh(2\pi\gamma/\omega_s )-\cos(2\pi(\omega-\omega_0)/\omega_s )} \ .
\end{equation}
The power spectrum is governed by the same formula as the transition
rate in \eqref{Wlargexx}, with a reinterpretation of the parameters. Therefore it exhibits similar features.
It has
an oscillating behavior in the parameter $\chi $, but at the special frequencies
\begin{equation}
    \omega-\omega_0 = \omega_s(n+\frac12 )\ ,
\end{equation}
the oscillating behavior is suppressed by an additional factor $1/\chi$, as $\sin (2\chi)$ has vanishing coefficient in this case.
As a function of $\omega$, the asymptotic power spectrum is periodic
with period $2/\omega_s$,
but the period  becomes $1/\omega_s$  at the special values of $\chi= n\pi/2$ where $\sin(2\chi)=0$.
On the other hand, at the resonant frequencies $\omega-\omega_0 =-n \omega_s $, the power spectrum undergoes the expected enhancement.

\begin{figure}[h!]
 \centering
\includegraphics[width=0.5\textwidth]{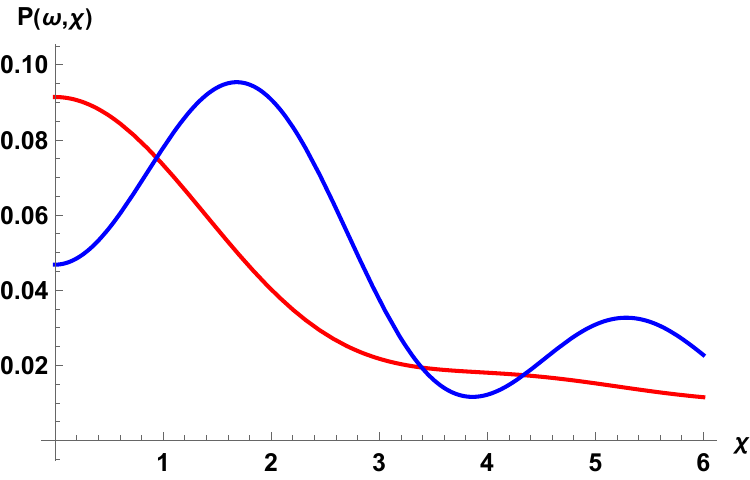}
 \caption
 {The exact $P$ given in \eqref{aaff} (in units of $1/{\rm GHz}^2$) as  a function of the dimensionless modulation index $\chi $. The oscillations are strongly suppressed
 at half-integer values of $(\omega-\omega_0)/\omega_s$. Here $\gamma=0.25$ GHz, $(\omega-\omega_0)=-\tfrac12 \omega_s$ (red) and $(\omega-\omega_0)=-0.7 \omega_s$ (blue).}
 \label{figexp}
 \end{figure}

The experimental values in \cite{metcalfe2010resolved} 
 for the modulation index $\chi $ are in the range $\chi \approx 0-2$ (see fig. 3b in \cite{metcalfe2010resolved}).
The driving frequency is $\omega_s/2\pi=\nu_s=1.05$ GHz and the linewidth is
250 MHz. 
It is instructive to see the changes in the fluorescence spectrum 
for these typical experimental values as one varies
the driving frequency.
Consider a resonance peak arising at $\omega-\omega_0=-n_0\omega_s$, for some integer $n_0$.
Taking,  for example, $n_0=1$, if  the sum is approximated by keeping only the terms $n=n_0, n_0\pm 1$, as compared to the exact formula, the power spectrum around this peak changes between 5 and 20 per cent, with bigger differences for larger $\chi$.
On the other hand, one can also observe the suppression of oscillations in $\chi $ occurring at half integer values of $(\omega-\omega_0)/\omega_s$ (corresponding to the minima
in fig. 3a in \cite{metcalfe2010resolved}).
By setting, for example, $\omega-\omega_0=-\tfrac12 \omega_s$,
$\omega_s/2\pi=\nu_s=1.05$ GHz, then the oscillations in the interval $\chi=1-6$
are significantly suppressed as compared with 
the oscillations at a generic value of $\omega-\omega_0$ (see fig. \ref{figexp}).
This illustrates the remarkable interference effect predicted by the exact analytic formula.

\medskip

\medskip

The study of quantum dots naturally prompts consideration of an area where the summation formula \eqref{general} can have multiple applications, namely cavity optomechanics \cite{aspelmeyer2014cavity}. For example, it is possible to have sideband cooling of a nanomechanical resonator with an embedded QD \cite{wilson2004laser}, which involves the quantized transfer of energy from a mechanical mode of the resonator to the applied optical field. This was studied with resonant spectroscopy in \cite{metcalfe2010resolved}.

The coupling of the resonant laser to the two-level system is described by adding to the Hamiltonian an additional interaction term
\begin{equation}
H_{\rm int}=-dE_0\cos(\omega_L t)\sigma_x,
\end{equation}
describing the dipole coupling of the QD to a laser field $E_0\cos(\omega_L t)$.
In \cite{metcalfe2010resolved}, the power spectrum of the fluorescence for weak excitations 
is found to be proportional to
\begin{equation}
P[\omega]=\sum^{\infty}_{\ell =-\infty}\left|\sum_{n=-\infty}^{\infty}\frac{J_{\ell+n}(\chi) J_n(\chi)}{\gamma-i(\omega_L-\omega_0+n\omega_s)}\right|^2 \delta(\omega-\omega_L+\ell \omega_s),
\label{eqn: resPspect}
\end{equation} which can be obtained by calculating the time dependence of the atomic dipole moment in the steady state.
This is of the form of a series of discrete lines at frequencies $\omega_\ell=\omega_L-\ell \omega_s$, spectrally separated
from the excitation frequency by multiples of the SAW frequency. We now give an analytical evaluation of the strength of these lines. 

The sum in \eqref{eqn: resPspect}  involves a novel application of the summation formula \eqref{general}, where now $\alpha$ or $\beta$ may be different from zero. 
This structure will be ubiquitous in cavity optomechanics.
The emission frequencies differing from the excitation frequency corresponds to the transfer of mechanical energy to the light field. 

Using \eqref{general2}, we find
\begin{eqnarray}
&&\sum_{n=-\infty }^{\infty }\frac{J_{\ell +
n}\left( z\right) J_{ n}\left( z\right) }{n+\mu }=\frac{\pi }{\sin \left( \pi \mu \right) }J_{\ell - \mu }\left( z\right) J_{ \mu }\left( z\right) , \qquad \ell\geq 0\ , 
\nonumber\\
&&\sum_{n=-\infty }^{\infty }\frac{J_{\ell +
n}\left( z\right) J_{ n}\left( z\right) }{n+\mu }=(-1)^{\ell }\, \frac{\pi }{\sin \left( \pi \mu \right) }J_{\mu-\ell }\left( z\right) J_{- \mu }\left( z\right) , \qquad \ell< 0\ .
\label{ffrq}
\end{eqnarray}
Therefore we find    
%
\begin{equation}
P[\omega]=P_+[\omega]+P_-[\omega]\ ,
\end{equation}
where $P_+$, $P_-$ refer to the contributions from
positive and negative $\ell$. Respectively, they describe the two sets of frequencies of the spectrum $\omega=\omega_s\mp \ell\omega_s
$. Explicitly,
\be
P_+[\omega]=\frac{1}{\omega_s^2}\left|\frac{\pi }{%
\sin \left( \pi \mu \right) }J_{ \mu }\left( \chi\right)   \right|^2
\sum^{\infty}_{\ell =0}\left|J_{\ell -\mu }\left( \chi \right)   \right|^2 \delta(\omega-\omega_L+\ell \omega_s)\ , 
\label{pkk}
\ee
\be
P_-[\omega]=\frac{1}{\omega_s^2}\left|\frac{\pi }{%
\sin \left( \pi \mu \right) }J_{- \mu }\left( \chi\right)   \right|^2
\sum^{\infty}_{\ell =1}\left|J_{\ell +\mu }\left( \chi \right)   \right|^2 \delta(\omega-\omega_L-\ell \omega_s)\ , 
\label{pkk2}
\ee
\begin{equation}
    \mu=\zeta+i\eta\ ,\qquad \zeta\equiv \frac{\omega_L-\omega_0}{\omega_s}\ ,\qquad \eta\equiv \frac{\gamma}{\omega_s}\ .
\end{equation}
$P_+[\omega]$ and $P_-[\omega]$  exhibit an oscillatory behavior in $\chi $ and in $\zeta$. 


For large $\chi $, the Bessel functions can be replaced by their asymptotic expressions.
Explicit formulas are given in appendix B. One can characterize the detailed features
   of the power spectrum and read frequencies and amplitudes thereof in terms of the physical parameters.
The power spectrum has different properties 
in the ``even" sector  $\omega=\omega_L-2k\omega_s$
and in the ``odd" sector  $\omega=\omega_L-(2k+1)\omega_s$.
 In particular, for $\omega=\omega_L-2k\omega_s$, $k\in\mathbb{Z}$, in the limit that the line width $\gamma\to 0$, the power spectrum takes the simple form
   \be
   \label{rruu}
P[\omega]_{\rm even}= 
     \frac{ \big(\cos(2\pi\zeta) +\sin(2\chi)\big)^2 }{\chi^2\omega_s^2\  \sin^2(\pi\zeta)} \sum_k \delta(\omega-\omega_L+2k\omega_s)\ .
   \ee
The poles at integer $\zeta $ correspond to 
  resonance peaks in the complete formula with $\gamma\neq 0$ (see appendix B). The height of the peak is proportional to $\big(1+\sin(2\chi )\big)^2/\chi^2$, exhibiting a  $1/\chi^2$ decrease at
 large $\chi$.
 Remarkably, for the special values 
 $$
 \chi =-\frac{\pi}{4}+\pi n\ ,\qquad n\in \mathbb{Z}\ ,
 $$  
 all resonance peaks in the detuning parameter $\zeta $ disappear  (modulo $O(1/\chi^3)$ corrections).
On the other hand, at the same special values of $\chi$, 
$P[\omega]_{\rm odd}$ becomes $O(1)$ instead of $O(1/\gamma^2)$ at the resonance peaks located at integer values of $\zeta $.

Another  feature is the fact that the power spectrum \eqref{rruu} identically vanishes 
at parameters satisfying  $\sin(2\chi) =-\cos(2\pi\zeta)$ (more precisely, at these special values $ P_{\rm even}[\omega]$ is strongly suppressed
and becomes $O(1/\chi^3)$). Similar features are shared by $ P_{\rm odd}[\omega]$.

It should be possible to test these intriguing properties in a laboratory setting.


\subsection{Power spectrum in other QD systems}

The use of Eq. \eqref{general2} is also applicable to a host of fluorescence spectra. The power spectrum is usually left in summation form (see for example \cite{blind1980resonance,ficek2001fluorescence,kryuchkyan2017resonance}). 
However, it is now possible to compute all the infinite sums and provide the corresponding analytical expressions. Just as in the earlier examples, the closed formulas reveal important aspects of the physics that become  manifest after resummation.





\subsubsection{Coherent scattered light and atomic inversion}  

The spectrum of light scattered by a two-level atom in an intense laser beam with
amplitude modulated light was considered long ago and just a few years after foundational works, such as \cite{mollow1969power} and \cite{cohen1977frontiers}. Analytical expressions already appeared in \cite{blind1980resonance}, in summation form, of a similar type as Eq. (\ref{eqn: resPspect}). 

In \cite{blind1980resonance} it was studied analytically the fluorescence spectrum in an amplitude modulated field,
composed of a strong resonant component of frequency $\omega _{0}$ and two considerably weaker sidebands of frequencies $\omega _{0}\pm \omega _{1}$. It was found that the spectrum is characterized by a central component, insensitive to the presence of the modulating fields, and a series of sidebands centered about the Rabi frequency $\Omega _{0}$ of the resonant field. The sidebands are located at frequencies $\Omega _{0}+\omega _{1}n$ for $n\in \mathbb{Z}$. 

For the steady state solution $\left\langle
\sigma _{+}(t)\right\rangle $ of the optical Bloch equations, the  spectral distribution of the
coherently scattered light  is  \cite{blind1980resonance} 
\[
I_{\rm coh}\left( \omega \right) \approx \frac{\pi \gamma ^{2}}{8} \sum_{k=-\infty
}^{\infty }Y_{k}Y_{k}^{\ast }\delta \left( \omega -\omega _{L}+k\omega
_{1}\right) ,
\]%
where 
\be
Y_{k}\equiv \sum_{m=-\infty }^{\infty }J_{m}\left( \frac{a\Omega _{0}}{\omega _{1}}%
\right) \left( \frac{J_{m-k}\left( \frac{a\Omega _{0}}{\omega _{1}}\right) }{%
\frac{3}{4}\gamma -i(\Omega _{0}+m\omega _{1})}-\frac{J_{m+k}\left( \frac{%
a\Omega _{0}}{\omega _{1}}\right) }{\frac{3}{4}\gamma +i(\Omega _{0}+m\omega
_{1})}\right) .
\label{uuh}
\ee
Using \eqref{general2}, we obtain
\[
Y_{k}=\frac{i\pi }{\omega _{1}}
\left( (-1)^{k}\frac{J_{-\rho} J_{k+\rho}}{\sin \left( \pi \rho \right)}+\frac{J_{\rho^*}J_{k-\rho^*}}
{\sin \left( \pi \rho^*\right) }\right) ,\qquad {\rm for\ } k\geq 0\ ,
\]%

\[
Y_{k}=\frac{i\pi }{\omega _{1}}\left( \frac{J_{\rho }J_{-k-\rho }}{
\sin \left( \pi \rho \right) }
+(-1)^{k}\frac{J_{-\rho^*}J_{-k+\rho^*}}{\sin \left( \pi \rho ^* \right) }\right)  ,\qquad {\rm for\ } k\leq 0\ ,
\]%
where
\begin{equation}
    \rho  \equiv\frac{\Omega _{0}}{\omega _{1}}+i\frac{3\gamma }{4\omega _{1}}\ .
\end{equation}
Therefore%
\[
I_{\rm coh}(\omega )\approx \frac{\pi^3 \gamma ^{2}}{8\omega_1^2}\left( \sum\limits_{k=0}^{\infty
}a_{k}\delta \left( \omega -\omega _{L}+k\omega _{1}\right)
+\sum^{-1}_{k=-\infty }b_{k}\delta \left( \omega -\omega _{L}+k\omega
_{1}\right)\right) ,
\]%
with 
\begin{eqnarray}
    a_{k}&=&\left| (-1)^{k}
    \frac{J_{-\rho}J_{k+\rho }}{\sin \left( \pi \rho \right) }+
    \frac{J_{\rho^*}J_{k-\rho^*}}{\sin \left( \pi \rho^*\right) }\right|^2 ,
\nonumber\\
b_{k} &=& 
\left|  \frac{J_{\rho }J_{-k-\rho }}{
\sin \left( \pi \rho \right) }
+(-1)^{k}\frac{J_{-\rho^*}J_{-k+\rho^*}}{\sin \left( \pi \rho ^* \right) }\right|^2  .\nonumber
\end{eqnarray}

\smallskip

\smallskip

\noindent Here the argument of all Bessel functions is $\frac{a\Omega _{0}}{\omega _{1}}$ as in the original expression \eqref{uuh}.
The formula now displays the explicit dependence of the spectral distribution in all physical parameters. It can be analysed in the same form
as done in previous examples.

\subsubsection{Steady state solutions. Atomic inversion}

The steady-state atomic dipole moment $\left\langle \sigma
_{+}(t)\right\rangle $ and inversion $\left\langle \sigma _{z}(t)\right\rangle $ are time-dependent and contain oscillations for all the harmonics of the modulation frequency $\omega _{1}$. The expectation value of $\left\langle \sigma _{z}(t)\right\rangle $ is equal to the difference between the upper-state and lower-state populations (atomic inversion), $\gamma $ denotes the spontaneous decay rate of the upper level \cite{blind1980resonance}. 

One can obtain approximate solutions of the optical Bloch equations, valid for strong driving fields at resonance ($\Omega \left( t\right) \gg \gamma ^{2}/16$). The result is \cite{blind1980resonance,thomann1980optical}
\[
z(t)=\left\langle \sigma _{z}(t)\right\rangle =-\gamma {\rm Re}\left(
\sum\limits_{m,k=-\infty }^{\infty }\frac{J_{m-k}\left( \frac{a\Omega _{0}}{%
\omega _{1}}\right) J_{m}\left( \frac{a\Omega _{0}}{\omega _{1}}\right) \exp
\left( -ik\omega _{1}t\right) }{\frac{3}{4}\gamma -i(\Omega _{0}+m\omega
_{1})}\right) 
\]%
and $y(t)=-2i\left\langle \sigma _{+}(t)\right\rangle $ is given by
a similar double sum expression,  with $-\gamma {\rm Re}\rightarrow
\gamma {\rm Im}$ \cite{blind1980resonance}. 

Thus, using \eqref{general2}, one can now evaluate the atomic inversion term $z_{0}$
(corresponding to $k=0$) and also any other subharmonic, for either 
positive or negative $k$. We obtain the following exact formulas for the  harmonics:
\be
z(t)=\sum_{k=-\infty}^\infty \beta_k\ e^{-ik\omega_1 t}
\ee
\bea
&&\beta_k= (-1)^{k}\,\frac{ \pi \gamma }{\omega _{1}}\ {\rm Im}\left( \frac{ J_{-\rho}J_{k+\rho }}{\sin \left( \pi \rho \right) }\right) \ , \quad {\rm for}\ k\geq 0  ,
\nonumber\\
&&\beta_k=\frac{\pi \gamma }{\omega _{1}}\ {\rm Im}\left( \frac{J_{\rho}J_{-k-\rho }}{\sin \left( \pi \rho \right) }\right)\ ,\ \ \ \qquad {\rm for}\ k<0 \ .
\eea
The harmonics satisfy the following recurrence relations, inherited from the
familiar recurrence relations of the Bessel functions,
\be
\beta_{k+1}+\beta_{k-1}=-\frac{2\omega_1(k+\rho)}{a\Omega_0}\, \beta_k\ ,\qquad k\neq 0\ .
\ee

\subsection{Asymmetric quantum dots with bi-chromatic electromagnetic field}

In \cite{kryuchkyan2017resonance}, an asymmetric quantum dot with broken inversion symmetry along the $z$ axis 
together with a bi-chromatic field was considered,
\be
{\bf E}(t)= {\bf E_1}\cos\omega_1 t +  {\bf E_2}\cos\omega_2 t\ ,
\ee
with ${\bf E_1}=(0,0,E_1)$, ${\bf E_2}=(E_2,0,0)$.
It was also assumed that the second frequency $\omega_2$ was close to the electronic resonance frequency, while the first frequency $\omega_1$ was far from resonance. The resonance condition is
\begin{equation}
\omega _{0}\pm \omega _{2}=n\omega _{1},  \label{res1}
\end{equation}%
and in \cite{kryuchkyan2017resonance} all modes except the resonant one are neglected. Therefore, the parameter $n$ above, denoting the number of modes, will appear in the power spectrum below. Then \begin{equation}
    F_{n}=-({{d}_{12}{E}_{2}}/{2\hbar })J_{n}\left( \widetilde{\omega }/\omega _{1}\right) 
\end{equation}
are the
Rabi frequencies of the considered system and 
\begin{equation}
\widetilde{\omega }=\frac{E_{1}(d_{22}-d_{11})}{\hbar }  \label{oo}
\end{equation}%
is the effective frequency, whereas $\varphi _{n}=\omega _{0}\pm \omega
_{2}-n\omega _{1}$ is the resonance detuning. The amplitudes of the fields are $E_{1}$ and $E_{2}$ and ${d}_{11}=\langle 1|ez|1\rangle$, ${d}_{22}=\langle
2|ez|2\rangle$ and ${d}_{12}={d}_{21}=\langle 1|ex|2\rangle$ are
the matrix elements of the operator of electric dipole moment
along the $z,x$ axes, and $e$ is the electron charge.

The problem in \cite{kryuchkyan2017resonance} is thus reduced 
to the effective two-level system driven by a monochromatic field with the
combined frequency $\varphi _{n}$. For simplicity in the formulas, one also defines $\Omega _{n}=\sqrt{\frac{%
\varphi _{n}^{2}}{4}+F_{n}^{2}}.$ 

Finally, using the
dressed-atom method to discuss resonant fluorescence, one finds for  the width of the transitions
\cite{kryuchkyan2017resonance} 
\begin{equation}
\Gamma_{11}= \frac{\gamma}{2}\left(
1+\frac{\varphi_n^2}{4\Omega_n^2} \right),\,\,\,\,
\Gamma_{12}=\frac{\gamma}{4}\left(
3-\frac{\varphi_n^2}{4\Omega_n^2} \right).
\end{equation}  
where $\gamma $ denotes the spontaneous emission rate.

The resulting inelastic power spectrum is given in terms of  an infinite set of Mollow triplets \cite{mollow1969power} ($x\equiv \widetilde{\omega }/\omega _{1}$):
\begin{eqnarray}
S_{2}(\omega ) &\sim &\frac{d_{12}^{2}}{4\pi }\bigg[ \left( 1-\Delta
_{S}^{2}\right) \left( \frac{d_{12}E_{2}}{2\hbar \Omega _{n}}\right)
^{2}J_{n}^{2}\left( x\right) \sum\limits_{m}\frac{J_{m}^{2}\left( {x}\right)
\Gamma _{11}}{[\omega -(n-m)\omega _{1}-\omega _{2}]^{2}+\Gamma _{11}^{2}}
\nonumber\\
&&+\frac{1}{2}\frac{\left( 1-{\varphi _{n}^{2}}/{4\Omega _{n}^{2}}\right)
^{2}}{\left( 1+{\varphi _{n}^{2}}/{4\Omega _{n}^{2}}\right) }\sum\limits_{m}%
\bigg( \frac{J_{m}^{2}\left( {x}\right) \Gamma _{12}}{[\omega -(n-m)\omega
_{1}-\omega _{2}+2\Omega _{n}]^{2}+\Gamma _{12}^{2}}  
\nonumber\\
&&+
\frac{J_{m}^{2}\left( {x}\right) \Gamma _{12}}{[\omega -(n-m)\omega_{1}-\omega _{2}-2\Omega _{n}]^{2}+\Gamma _{12}^{2}}\bigg)\bigg].
\label{gryy}
\end{eqnarray}%
The infinite  sums can now be carried out using  our summation formula \eqref{Wexacta}, giving 
\begin{eqnarray}
S_{2}(\omega ) &\sim &-\frac{d_{12}^{2}}{4\pi \omega _{1}}\bigg[\left( 1-\Delta
_{S}^{2}\right) \left( \frac{d_{12}E_{2}}{2\hbar \Omega _{n}}\right)
^{2}J_{n}^{2}\left( x\right) {\rm Im}\left( \frac{\pi }{\sin \left( \pi \nu
_{1}\right) }J_{\nu _{1}}\left( {x}\right) J_{-\nu _{1}}\left( {x}\right)
\right) 
\nonumber\\
&&+\frac{\left( 1-{\varphi _{n}^{2}}/{4\Omega _{n}^{2}}%
\right) ^{2}}{2\left( 1+{\varphi _{n}^{2}}/{4\Omega _{n}^{2}}\right) }{\rm Im%
}\left( \frac{\pi }{\sin \left( \pi \nu _{2}\right) }J_{\nu _{2}}\left( {x}%
\right) J_{-\nu _{2}}\left( {x}\right) + \frac{\pi }{\sin \left( \pi \nu _{3}\right) }J_{\nu _{3}}\left( {x}%
\right) J_{-\nu _{3}}\left( {x}\right) \right) .
\end{eqnarray}
where the spectral parameters of the Bessel functions are:            
\begin{eqnarray*}
\nu _{1} &=& n+\frac{1}{\omega _{1}}\left( \omega_2 -\omega
+i\Gamma _{11}\right) , \\
\nu _{2} &=& n+ \frac{1}{\omega _{1}}\left( \omega_2 -\omega -2\Omega _{n}+i\Gamma _{12}\right) ,  \\
\nu _{3} &=& n+ \frac{1}{\omega _{1}}\left( \omega_2 -\omega +2\Omega _{n}+i\Gamma _{12}\right)  .
\end{eqnarray*}%
%
%

\section{Concluding remarks}

Sideband multiphoton spectra similar to 
   \eqref{uno}, \eqref{eqn: resPspect}, \eqref{gryy} are ubiquitous in 
 two-level systems with harmonic modulation.
 In this paper we provided exact formulas that compute the infinite sums, enabling exploration of different spectra across all parameter space.

In Section II we have seen that the transition rate is an oscillating function of the driving field amplitude 
that never vanishes. 
At large amplitude the exact transition rate can be written in terms of 
trigonometric functions, providing simple expressions to test a number of features.
In particular, one interesting feature is the suppression of these oscillations for
half-integer values of the detuning.

As a function of the detuning parameter -- represented by $\mathcal{E}$ in section II -- the transition rate
exhibits the expected resonance peaks at integer $\mathcal{E}$.
We also noticed  a significant change of behavior  when $x$ is an integer multiple of $\pi/2$, where the shape of the spectrum simplifies and the period is halved.
We have also analytically determined the minimum and maximum values for each period of oscillations in \eqref{vvnn}.

Analogous properties were identified in the absorption spectra of quantum dots  in section III, where we also considered coupling to 
a more general configuration of driving fields. This includes
either the addition of an extra field or the consideration of bichromatic
fields. In these cases, the required summation formula  appears in a different form, exploiting other particular instances of the analytical expression \eqref{general2}, beyond the one employed in section II. In all cases, closed formulas for the various spectra are obtained, which display explicit dependence on all physical parameters.
Clearly, it would be worthwhile to conduct experiments to test the properties predicted by the analytical formulas in  superconducting qubits or quantum dot devices. 

\section*{Acknowledgements}

We thank David Pérez-García and Sergio Valenzuela for  useful comments.
J.G.R. acknowledges financial support from grants 2021-SGR-249 (Generalitat de Catalunya) and MINECO  PID2019-105614GB-C21. M.T. acknowledges financial support from FEI-EU-22-06, funded by Universidad Complutense de Madrid, and grant PID2020-113523GB-I00, funded by the Spanish Ministry of Science and Innovation. This work is financially supported by the  Ministry of Economic Affairs and Digital Transformation of the Spanish Government through the QUANTUM ENIA project call - Quantum Spain project, by the European Union through the Recovery, Transformation and Resilience Plan - NextGenerationEU within the framework of the Digital Spain 2026 Agenda.

\appendix

\section{The summation formula}

The more general form of the summation formula, found in 
\cite{newberger1982new}, is 
\begin{equation}
\sum_{n=-\infty }^{\infty }\frac{\left( -1\right) ^{n}J_{\alpha +\gamma
n}\left( z\right) J_{\beta -\gamma n}\left( x\right) }{n+\mu }=\frac{\pi }{%
\sin \left( \pi \mu \right) }J_{\alpha -\gamma \mu }\left( x\right) J_{\beta
+\gamma \mu }\left( x\right) ,  \label{general}
\end{equation}%
where $\mu \in \mathbb{C}/\mathbb{Z}
$, $\alpha ,\beta ,x\in 
\mathbb{C}$,
$\gamma \in \left( 0,1\right]$, and $\Re\left( \alpha +\beta \right) >-1$. 
It is useful to  explicitly quote the particular case where $\gamma =1$, and $\alpha=p,\ \beta=q $, with $p,\ q \in\mathbb{Z}$, since this case  frequently arises in various fluorescence spectra.
The formula takes the form
\begin{equation}
\boxed{
\sum_{n=-\infty }^{\infty }\frac{J_{n+p}\left( z\right) J_{n-q}\left( x\right) }{n+\mu }=(-1)^q \frac{\pi }{%
\sin \left( \pi \mu \right) }J_{p - \mu }\left( x\right) J_{q
+ \mu }\left( x\right) }  \label{general2}
\end{equation}%
where we used  $J_{-n}(x)=(-1)^nJ_n(x)$ and convergence now requires 
$p+q>-1$.

Here we provide a simple proof in the case $\alpha =\beta =0$ and $\gamma =1$, referring to \cite{newberger1982new} for the general case. 
Interestingly, the product of
two Bessel functions admit different integral
representations \cite{watson1922treatise} in terms of the integration of a single Bessel function and
an additional simple trigonometric or hyperbolic function. One of these
representations is called of Nicholson type,%
\begin{equation}
J_{n}\left( x\right) J_{-n}\left( x\right) =\frac{2}{\pi }
\int_{0}^{\frac{\pi}{2}} J_{0}\left( 2x\cos \theta \right) \cos (2n\theta )d\theta \ .
\label{integral rep}
\end{equation}%
The Bessel function in the integrand is now independent of the summation index. Therefore, the summation
only involves the cos term. The simple nature of the formula and of its
derivation  lies in the fact that the result of the summation%
\begin{equation}
\sum\limits_{n=-\infty }^{\infty }\frac{\left( -1\right) ^{n}\cos n\phi }{%
n+\mu }=\frac{\pi \cos \mu \phi }{\sin \pi \mu }\ ,  \label{sum}
\end{equation}%
where $\phi \in \left[ -\pi ,\pi \right] ,$ which is satisfied in our case,
does not modify the analytical form of the integrand, because the r.h.s. of (\ref{sum}) also has a cosine with the integration variable. Therefore, the
resulting integral expression after the summation has the same integral
representation of the product of two Bessel functions \footnote{%
The integral representation (\ref{integral rep}) holds the
same way for more general indices, including complex.}. Therefore,
\begin{eqnarray}
\sum_{n=-\infty }^{\infty }\frac{J_{n}\left( x\right)
J_{n}\left( x\right) }{n+\mu } &=&\frac{2}{\sin \pi \mu }\int_{0}^{\pi
/2}J_{0}\left( 2x\cos \theta \right) \cos (2\mu \theta )d\theta \nonumber \\
&=&\frac{\pi J_{\mu}\left( x\right) J_{-\mu}\left( x\right) }{\sin \pi \mu }\ .
\label{neuber}
\end{eqnarray}
One of the first appearances, with a proof, of this formula is in \cite{hall1965ion}. The very first reference we could trace is actually \cite{sen1952solar}.   This formula  was  rediscovered by Newberger \cite{newberger1982new}, who also generalized it
in the form \eqref{general}.

\section{Asymptotic formulas for fluorescence spectrum}

For large $\chi$, one can again use the asymptotic formula for the Bessel function, giving
\begin{equation}
  P[\omega]\approx  P[\omega]_{\rm asym}=P[\omega]_{\rm even}+P_+[\omega]_{\rm odd}+P_-[\omega]_{\rm odd}\ ,
    \end{equation}
with    
\begin{eqnarray}    
   &&P[\omega]_{\rm even}= 
    \frac{4}{\chi^2\omega_s^2}\frac{ \big|\cos(\chi-\mu\frac{\pi}{2}-\frac{\pi}{4})\big|^2 \big|\cos(\chi+\mu\frac{\pi}{2}-\frac{\pi}{4})\big|^2}{|\sin(\pi\mu)|^2} \sum_k \delta(\omega-\omega_L+2k\omega_s)\ ,
    \nonumber\\
    &&P_+[\omega]_{\rm odd}= \frac{4}{\chi^2\omega_s^2}\frac{ \big|\cos(\chi-\mu\frac{\pi}{2}-\frac{\pi}{4})\big|^2 \big|\sin(\chi+\mu\frac{\pi}{2}-\frac{\pi}{4})\big|^2}{|\sin(\pi\mu)|^2} \sum_{k=0}^\infty \delta(\omega-\omega_L+(2k+1) \omega_s)\ ,
     \nonumber\\
    &&P_-[\omega]_{\rm odd}= \frac{4}{\chi^2\omega_s^2}\frac{ \big|\cos(\chi+\mu\frac{\pi}{2}-\frac{\pi}{4})\big|^2 \big|\sin(\chi-\mu\frac{\pi}{2}-\frac{\pi}{4})\big|^2}{|\sin(\pi\mu)|^2} \sum_{k=0}^\infty \delta(\omega-\omega_L-(2k+1) \omega_s)\ .
  \nonumber
\end{eqnarray}
Computing the modulus, we find
\begin{eqnarray}    
   &&P[\omega]_{\rm even}= 
     \frac{ \big(\cosh (2 \pi  \eta )\cos(2\pi\zeta) +\sin(2\chi)\big)^2 +\sin^2 (2 \pi  \zeta ) \sinh^2 (2 \pi  \eta )}{\chi^2\omega_s^2\ \big( \sin^2(\pi\zeta)+\sinh^2 ( \pi  \eta )\big)} \sum_k \delta(\omega-\omega_L+2k\omega_s)\ ,
    \nonumber\\
    &&P_+[\omega]_{\rm odd}=   \frac{ \big(\cosh (2 \pi  \eta )\sin(2\pi\zeta) -\cos(2\chi)\big)^2 +\cos^2 (2 \pi  \zeta ) \sinh^2 (2 \pi  \eta )}{\chi^2\omega_s^2 \ \big( \sin^2(\pi\zeta)+\sinh^2 ( \pi  \eta )\big)}\sum_{k=0}^\infty \delta(\omega-\omega_L+(2k+1) \omega_s),
    \nonumber\\
    &&P_-[\omega]_{\rm odd}=   \frac{ \big(\cosh (2 \pi  \eta )\sin(2\pi\zeta) +\cos(2\chi)\big)^2 +\cos^2 (2 \pi  \zeta ) \sinh^2 (2 \pi  \eta )}{\chi^2\omega_s^2 \ \big( \sin^2(\pi\zeta)+\sinh^2 ( \pi  \eta )\big)}\sum_{k=0}^\infty \delta(\omega-\omega_L-(2k+1) \omega_s).
    \nonumber
\end{eqnarray}

\bibliography{spin3}

\begin{thebibliography}{33}%
\makeatletter
\providecommand \@ifxundefined [1]{%
 \@ifx{#1\undefined}
}%
\providecommand \@ifnum [1]{%
 \ifnum #1\expandafter \@firstoftwo
 \else \expandafter \@secondoftwo
 \fi
}%
\providecommand \@ifx [1]{%
 \ifx #1\expandafter \@firstoftwo
 \else \expandafter \@secondoftwo
 \fi
}%
\providecommand \natexlab [1]{#1}%
\providecommand \enquote  [1]{``#1''}%
\providecommand \bibnamefont  [1]{#1}%
\providecommand \bibfnamefont [1]{#1}%
\providecommand \citenamefont [1]{#1}%
\providecommand \href@noop [0]{\@secondoftwo}%
\providecommand \href [0]{\begingroup \@sanitize@url \@href}%
\providecommand \@href[1]{\@@startlink{#1}\@@href}%
\providecommand \@@href[1]{\endgroup#1\@@endlink}%
\providecommand \@sanitize@url [0]{\catcode `\\12\catcode `\$12\catcode
  `\&12\catcode `\#12\catcode `\^12\catcode `\_12\catcode `\%12\relax}%
\providecommand \@@startlink[1]{}%
\providecommand \@@endlink[0]{}%
\providecommand \url  [0]{\begingroup\@sanitize@url \@url }%
\providecommand \@url [1]{\endgroup\@href {#1}{\urlprefix }}%
\providecommand \urlprefix  [0]{URL }%
\providecommand \Eprint [0]{\href }%
\providecommand \doibase [0]{https://doi.org/}%
\providecommand \selectlanguage [0]{\@gobble}%
\providecommand \bibinfo  [0]{\@secondoftwo}%
\providecommand \bibfield  [0]{\@secondoftwo}%
\providecommand \translation [1]{[#1]}%
\providecommand \BibitemOpen [0]{}%
\providecommand \bibitemStop [0]{}%
\providecommand \bibitemNoStop [0]{.\EOS\space}%
\providecommand \EOS [0]{\spacefactor3000\relax}%
\providecommand \BibitemShut  [1]{\csname bibitem#1\endcsname}%
\let\auto@bib@innerbib\@empty
\bibitem [{\citenamefont {Krantz}\ \emph {et~al.}(2019)\citenamefont {Krantz},
  \citenamefont {Kjaergaard}, \citenamefont {Yan}, \citenamefont {Orlando},
  \citenamefont {Gustavsson},\ and\ \citenamefont
  {Oliver}}]{krantz2019quantum}%
  \BibitemOpen
  \bibfield  {author} {\bibinfo {author} {\bibfnamefont {P.}~\bibnamefont
  {Krantz}}, \bibinfo {author} {\bibfnamefont {M.}~\bibnamefont {Kjaergaard}},
  \bibinfo {author} {\bibfnamefont {F.}~\bibnamefont {Yan}}, \bibinfo {author}
  {\bibfnamefont {T.~P.}\ \bibnamefont {Orlando}}, \bibinfo {author}
  {\bibfnamefont {S.}~\bibnamefont {Gustavsson}},\ and\ \bibinfo {author}
  {\bibfnamefont {W.~D.}\ \bibnamefont {Oliver}},\ }\bibfield  {title}
  {\bibinfo {title} {A quantum engineer's guide to superconducting qubits},\
  }\href@noop {} {\bibfield  {journal} {\bibinfo  {journal} {Applied physics
  reviews}\ }\textbf {\bibinfo {volume} {6}} (\bibinfo {year}
  {2019})}\BibitemShut {NoStop}%
\bibitem [{\citenamefont {Shevchenko}\ \emph {et~al.}(2010)\citenamefont
  {Shevchenko}, \citenamefont {Ashhab},\ and\ \citenamefont
  {Nori}}]{shevchenko2010landau}%
  \BibitemOpen
  \bibfield  {author} {\bibinfo {author} {\bibfnamefont {S.~N.}\ \bibnamefont
  {Shevchenko}}, \bibinfo {author} {\bibfnamefont {S.}~\bibnamefont {Ashhab}},\
  and\ \bibinfo {author} {\bibfnamefont {F.}~\bibnamefont {Nori}},\ }\bibfield
  {title} {\bibinfo {title} {Landau--zener--st{\"u}ckelberg interferometry},\
  }\href@noop {} {\bibfield  {journal} {\bibinfo  {journal} {Physics Reports}\
  }\textbf {\bibinfo {volume} {492}},\ \bibinfo {pages} {1} (\bibinfo {year}
  {2010})}\BibitemShut {NoStop}%
\bibitem [{\citenamefont {Ivakhnenko}\ \emph {et~al.}(2023)\citenamefont
  {Ivakhnenko}, \citenamefont {Shevchenko},\ and\ \citenamefont
  {Nori}}]{ivakhnenko2023nonadiabatic}%
  \BibitemOpen
  \bibfield  {author} {\bibinfo {author} {\bibfnamefont {V.}~\bibnamefont
  {Ivakhnenko}}, \bibinfo {author} {\bibfnamefont {S.~N.}\ \bibnamefont
  {Shevchenko}},\ and\ \bibinfo {author} {\bibfnamefont {F.}~\bibnamefont
  {Nori}},\ }\bibfield  {title} {\bibinfo {title} {Nonadiabatic
  landau--zener--st{\"u}ckelberg--majorana transitions, dynamics, and
  interference},\ }\href@noop {} {\bibfield  {journal} {\bibinfo  {journal}
  {Physics Reports}\ }\textbf {\bibinfo {volume} {995}},\ \bibinfo {pages} {1}
  (\bibinfo {year} {2023})}\BibitemShut {NoStop}%
\bibitem [{\citenamefont {Di~Giacomo}\ and\ \citenamefont
  {Nikitin}(2005)}]{di2005majorana}%
  \BibitemOpen
  \bibfield  {author} {\bibinfo {author} {\bibfnamefont {F.}~\bibnamefont
  {Di~Giacomo}}\ and\ \bibinfo {author} {\bibfnamefont {E.~E.}\ \bibnamefont
  {Nikitin}},\ }\bibfield  {title} {\bibinfo {title} {The majorana formula and
  the landau--zener--st{\"u}ckelberg treatment of the avoided crossing
  problem},\ }\href@noop {} {\bibfield  {journal} {\bibinfo  {journal}
  {Physics-Uspekhi}\ }\textbf {\bibinfo {volume} {48}},\ \bibinfo {pages} {515}
  (\bibinfo {year} {2005})}\BibitemShut {NoStop}%
\bibitem [{\citenamefont {Kofman}\ \emph {et~al.}(2023)\citenamefont {Kofman},
  \citenamefont {Ivakhnenko}, \citenamefont {Shevchenko},\ and\ \citenamefont
  {Nori}}]{kofman2023majorana}%
  \BibitemOpen
  \bibfield  {author} {\bibinfo {author} {\bibfnamefont {P.~O.}\ \bibnamefont
  {Kofman}}, \bibinfo {author} {\bibfnamefont {O.~V.}\ \bibnamefont
  {Ivakhnenko}}, \bibinfo {author} {\bibfnamefont {S.~N.}\ \bibnamefont
  {Shevchenko}},\ and\ \bibinfo {author} {\bibfnamefont {F.}~\bibnamefont
  {Nori}},\ }\bibfield  {title} {\bibinfo {title} {Majorana’s approach to
  nonadiabatic transitions validates the adiabatic-impulse approximation},\
  }\href@noop {} {\bibfield  {journal} {\bibinfo  {journal} {Scientific
  Reports}\ }\textbf {\bibinfo {volume} {13}},\ \bibinfo {pages} {5053}
  (\bibinfo {year} {2023})}\BibitemShut {NoStop}%
\bibitem [{\citenamefont {Nikitin}(1999)}]{nikitin1999nonadiabatic}%
  \BibitemOpen
  \bibfield  {author} {\bibinfo {author} {\bibfnamefont {E.}~\bibnamefont
  {Nikitin}},\ }\bibfield  {title} {\bibinfo {title} {Nonadiabatic transitions:
  What we learned from old masters and how much we owe them},\ }\href@noop {}
  {\bibfield  {journal} {\bibinfo  {journal} {Annual review of physical
  chemistry}\ }\textbf {\bibinfo {volume} {50}},\ \bibinfo {pages} {1}
  (\bibinfo {year} {1999})}\BibitemShut {NoStop}%
\bibitem [{\citenamefont {Berns}\ \emph {et~al.}(2006)\citenamefont {Berns},
  \citenamefont {Oliver}, \citenamefont {Valenzuela}, \citenamefont {Shytov},
  \citenamefont {Berggren}, \citenamefont {Levitov},\ and\ \citenamefont
  {Orlando}}]{berns2006coherent}%
  \BibitemOpen
  \bibfield  {author} {\bibinfo {author} {\bibfnamefont {D.}~\bibnamefont
  {Berns}}, \bibinfo {author} {\bibfnamefont {W.}~\bibnamefont {Oliver}},
  \bibinfo {author} {\bibfnamefont {S.}~\bibnamefont {Valenzuela}}, \bibinfo
  {author} {\bibfnamefont {A.}~\bibnamefont {Shytov}}, \bibinfo {author}
  {\bibfnamefont {K.}~\bibnamefont {Berggren}}, \bibinfo {author}
  {\bibfnamefont {L.}~\bibnamefont {Levitov}},\ and\ \bibinfo {author}
  {\bibfnamefont {T.}~\bibnamefont {Orlando}},\ }\bibfield  {title} {\bibinfo
  {title} {Coherent quasiclassical dynamics of a persistent current qubit},\
  }\href@noop {} {\bibfield  {journal} {\bibinfo  {journal} {Physical review
  letters}\ }\textbf {\bibinfo {volume} {97}},\ \bibinfo {pages} {150502}
  (\bibinfo {year} {2006})}\BibitemShut {NoStop}%
\bibitem [{\citenamefont {Oliver}\ \emph {et~al.}(2005)\citenamefont {Oliver},
  \citenamefont {Yu}, \citenamefont {Lee}, \citenamefont {Berggren},
  \citenamefont {Levitov},\ and\ \citenamefont {Orlando}}]{oliver2005mach}%
  \BibitemOpen
  \bibfield  {author} {\bibinfo {author} {\bibfnamefont {W.~D.}\ \bibnamefont
  {Oliver}}, \bibinfo {author} {\bibfnamefont {Y.}~\bibnamefont {Yu}}, \bibinfo
  {author} {\bibfnamefont {J.~C.}\ \bibnamefont {Lee}}, \bibinfo {author}
  {\bibfnamefont {K.~K.}\ \bibnamefont {Berggren}}, \bibinfo {author}
  {\bibfnamefont {L.~S.}\ \bibnamefont {Levitov}},\ and\ \bibinfo {author}
  {\bibfnamefont {T.~P.}\ \bibnamefont {Orlando}},\ }\bibfield  {title}
  {\bibinfo {title} {Mach-zehnder interferometry in a strongly driven
  superconducting qubit},\ }\href@noop {} {\bibfield  {journal} {\bibinfo
  {journal} {Science}\ }\textbf {\bibinfo {volume} {310}},\ \bibinfo {pages}
  {1653} (\bibinfo {year} {2005})}\BibitemShut {NoStop}%
\bibitem [{\citenamefont {Rudner}\ \emph {et~al.}(2008)\citenamefont {Rudner},
  \citenamefont {Shytov}, \citenamefont {Levitov}, \citenamefont {Berns},
  \citenamefont {Oliver}, \citenamefont {Valenzuela},\ and\ \citenamefont
  {Orlando}}]{rudner2008quantum}%
  \BibitemOpen
  \bibfield  {author} {\bibinfo {author} {\bibfnamefont {M.~S.}\ \bibnamefont
  {Rudner}}, \bibinfo {author} {\bibfnamefont {A.}~\bibnamefont {Shytov}},
  \bibinfo {author} {\bibfnamefont {L.~S.}\ \bibnamefont {Levitov}}, \bibinfo
  {author} {\bibfnamefont {D.~M.}\ \bibnamefont {Berns}}, \bibinfo {author}
  {\bibfnamefont {W.~D.}\ \bibnamefont {Oliver}}, \bibinfo {author}
  {\bibfnamefont {S.~O.}\ \bibnamefont {Valenzuela}},\ and\ \bibinfo {author}
  {\bibfnamefont {T.~P.}\ \bibnamefont {Orlando}},\ }\bibfield  {title}
  {\bibinfo {title} {Quantum phase tomography of a strongly driven qubit},\
  }\href@noop {} {\bibfield  {journal} {\bibinfo  {journal} {Physical review
  letters}\ }\textbf {\bibinfo {volume} {101}},\ \bibinfo {pages} {190502}
  (\bibinfo {year} {2008})}\BibitemShut {NoStop}%
\bibitem [{\citenamefont {Li}\ \emph {et~al.}(2013)\citenamefont {Li},
  \citenamefont {Silveri}, \citenamefont {Kumar}, \citenamefont {Pirkkalainen},
  \citenamefont {Veps{\"a}l{\"a}inen}, \citenamefont {Chien}, \citenamefont
  {Tuorila}, \citenamefont {Sillanp{\"a}{\"a}}, \citenamefont {Hakonen},
  \citenamefont {Thuneberg} \emph {et~al.}}]{li2013motional}%
  \BibitemOpen
  \bibfield  {author} {\bibinfo {author} {\bibfnamefont {J.}~\bibnamefont
  {Li}}, \bibinfo {author} {\bibfnamefont {M.}~\bibnamefont {Silveri}},
  \bibinfo {author} {\bibfnamefont {K.}~\bibnamefont {Kumar}}, \bibinfo
  {author} {\bibfnamefont {J.-M.}\ \bibnamefont {Pirkkalainen}}, \bibinfo
  {author} {\bibfnamefont {A.}~\bibnamefont {Veps{\"a}l{\"a}inen}}, \bibinfo
  {author} {\bibfnamefont {W.}~\bibnamefont {Chien}}, \bibinfo {author}
  {\bibfnamefont {J.}~\bibnamefont {Tuorila}}, \bibinfo {author} {\bibfnamefont
  {M.}~\bibnamefont {Sillanp{\"a}{\"a}}}, \bibinfo {author} {\bibfnamefont
  {P.}~\bibnamefont {Hakonen}}, \bibinfo {author} {\bibfnamefont
  {E.}~\bibnamefont {Thuneberg}}, \emph {et~al.},\ }\bibfield  {title}
  {\bibinfo {title} {Motional averaging in a superconducting qubit},\
  }\href@noop {} {\bibfield  {journal} {\bibinfo  {journal} {Nature
  communications}\ }\textbf {\bibinfo {volume} {4}},\ \bibinfo {pages} {1420}
  (\bibinfo {year} {2013})}\BibitemShut {NoStop}%
\bibitem [{\citenamefont {Gefen}\ \emph {et~al.}(1987)\citenamefont {Gefen},
  \citenamefont {Ben-Jacob},\ and\ \citenamefont {Caldeira}}]{gefen1987zener1}%
  \BibitemOpen
  \bibfield  {author} {\bibinfo {author} {\bibfnamefont {Y.}~\bibnamefont
  {Gefen}}, \bibinfo {author} {\bibfnamefont {E.}~\bibnamefont {Ben-Jacob}},\
  and\ \bibinfo {author} {\bibfnamefont {A.~O.}\ \bibnamefont {Caldeira}},\
  }\bibfield  {title} {\bibinfo {title} {Zener transitions in dissipative
  driven systems},\ }\href@noop {} {\bibfield  {journal} {\bibinfo  {journal}
  {Physical Review B}\ }\textbf {\bibinfo {volume} {36}},\ \bibinfo {pages}
  {2770} (\bibinfo {year} {1987})}\BibitemShut {NoStop}%
\bibitem [{\citenamefont {Shimshoni}\ and\ \citenamefont
  {Gefen}(1991)}]{shimshoni1991onset}%
  \BibitemOpen
  \bibfield  {author} {\bibinfo {author} {\bibfnamefont {E.}~\bibnamefont
  {Shimshoni}}\ and\ \bibinfo {author} {\bibfnamefont {Y.}~\bibnamefont
  {Gefen}},\ }\bibfield  {title} {\bibinfo {title} {Onset of dissipation in
  zener dynamics: relaxation versus dephasing},\ }\href@noop {} {\bibfield
  {journal} {\bibinfo  {journal} {Annals of Physics}\ }\textbf {\bibinfo
  {volume} {210}},\ \bibinfo {pages} {16} (\bibinfo {year} {1991})}\BibitemShut
  {NoStop}%
\bibitem [{\citenamefont {Gefen}\ and\ \citenamefont
  {Thouless}(1987)}]{gefen1987zener2}%
  \BibitemOpen
  \bibfield  {author} {\bibinfo {author} {\bibfnamefont {Y.}~\bibnamefont
  {Gefen}}\ and\ \bibinfo {author} {\bibfnamefont {D.}~\bibnamefont
  {Thouless}},\ }\bibfield  {title} {\bibinfo {title} {Zener transitions and
  energy dissipation in small driven systems},\ }\href@noop {} {\bibfield
  {journal} {\bibinfo  {journal} {Physical review letters}\ }\textbf {\bibinfo
  {volume} {59}},\ \bibinfo {pages} {1752} (\bibinfo {year}
  {1987})}\BibitemShut {NoStop}%
\bibitem [{\citenamefont {Metcalfe}\ \emph {et~al.}(2010)\citenamefont
  {Metcalfe}, \citenamefont {Carr}, \citenamefont {Muller}, \citenamefont
  {Solomon},\ and\ \citenamefont {Lawall}}]{metcalfe2010resolved}%
  \BibitemOpen
  \bibfield  {author} {\bibinfo {author} {\bibfnamefont {M.}~\bibnamefont
  {Metcalfe}}, \bibinfo {author} {\bibfnamefont {S.~M.}\ \bibnamefont {Carr}},
  \bibinfo {author} {\bibfnamefont {A.}~\bibnamefont {Muller}}, \bibinfo
  {author} {\bibfnamefont {G.~S.}\ \bibnamefont {Solomon}},\ and\ \bibinfo
  {author} {\bibfnamefont {J.}~\bibnamefont {Lawall}},\ }\bibfield  {title}
  {\bibinfo {title} {Resolved sideband emission of inas/gaas quantum dots
  strained by surface acoustic waves},\ }\href@noop {} {\bibfield  {journal}
  {\bibinfo  {journal} {Physical review letters}\ }\textbf {\bibinfo {volume}
  {105}},\ \bibinfo {pages} {037401} (\bibinfo {year} {2010})}\BibitemShut
  {NoStop}%
\bibitem [{\citenamefont {Blind}\ \emph {et~al.}(1980)\citenamefont {Blind},
  \citenamefont {Fontana},\ and\ \citenamefont {Thomann}}]{blind1980resonance}%
  \BibitemOpen
  \bibfield  {author} {\bibinfo {author} {\bibfnamefont {B.}~\bibnamefont
  {Blind}}, \bibinfo {author} {\bibfnamefont {P.}~\bibnamefont {Fontana}},\
  and\ \bibinfo {author} {\bibfnamefont {P.}~\bibnamefont {Thomann}},\
  }\bibfield  {title} {\bibinfo {title} {Resonance fluorescence spectrum of
  intense amplitude modulated laser light},\ }\href@noop {} {\bibfield
  {journal} {\bibinfo  {journal} {Journal of Physics B: Atomic and Molecular
  Physics}\ }\textbf {\bibinfo {volume} {13}},\ \bibinfo {pages} {2717}
  (\bibinfo {year} {1980})}\BibitemShut {NoStop}%
\bibitem [{\citenamefont {Kryuchkyan}\ \emph {et~al.}(2017)\citenamefont
  {Kryuchkyan}, \citenamefont {Shahnazaryan}, \citenamefont {Kibis},\ and\
  \citenamefont {Shelykh}}]{kryuchkyan2017resonance}%
  \BibitemOpen
  \bibfield  {author} {\bibinfo {author} {\bibfnamefont {G.~Y.}\ \bibnamefont
  {Kryuchkyan}}, \bibinfo {author} {\bibfnamefont {V.}~\bibnamefont
  {Shahnazaryan}}, \bibinfo {author} {\bibfnamefont {O.~V.}\ \bibnamefont
  {Kibis}},\ and\ \bibinfo {author} {\bibfnamefont {I.}~\bibnamefont
  {Shelykh}},\ }\bibfield  {title} {\bibinfo {title} {Resonance fluorescence
  from an asymmetric quantum dot dressed by a bichromatic electromagnetic
  field},\ }\href@noop {} {\bibfield  {journal} {\bibinfo  {journal} {Physical
  Review A}\ }\textbf {\bibinfo {volume} {95}},\ \bibinfo {pages} {013834}
  (\bibinfo {year} {2017})}\BibitemShut {NoStop}%
\bibitem [{\citenamefont {Mollow}(1969)}]{mollow1969power}%
  \BibitemOpen
  \bibfield  {author} {\bibinfo {author} {\bibfnamefont {B.}~\bibnamefont
  {Mollow}},\ }\bibfield  {title} {\bibinfo {title} {Power spectrum of light
  scattered by two-level systems},\ }\href@noop {} {\bibfield  {journal}
  {\bibinfo  {journal} {Physical Review}\ }\textbf {\bibinfo {volume} {188}}
  (\bibinfo {year} {1969})}\BibitemShut {NoStop}%
\bibitem [{\citenamefont {Otxoa}\ \emph {et~al.}(2019)\citenamefont {Otxoa},
  \citenamefont {Chatterjee}, \citenamefont {Shevchenko}, \citenamefont
  {Barraud}, \citenamefont {Nori},\ and\ \citenamefont
  {Gonzalez-Zalba}}]{otxoa2019quantum}%
  \BibitemOpen
  \bibfield  {author} {\bibinfo {author} {\bibfnamefont {R.~M.}\ \bibnamefont
  {Otxoa}}, \bibinfo {author} {\bibfnamefont {A.}~\bibnamefont {Chatterjee}},
  \bibinfo {author} {\bibfnamefont {S.~N.}\ \bibnamefont {Shevchenko}},
  \bibinfo {author} {\bibfnamefont {S.}~\bibnamefont {Barraud}}, \bibinfo
  {author} {\bibfnamefont {F.}~\bibnamefont {Nori}},\ and\ \bibinfo {author}
  {\bibfnamefont {M.~F.}\ \bibnamefont {Gonzalez-Zalba}},\ }\bibfield  {title}
  {\bibinfo {title} {Quantum interference capacitor based on double-passage
  landau-zener-st{\"u}ckelberg-majorana interferometry},\ }\href@noop {}
  {\bibfield  {journal} {\bibinfo  {journal} {Physical Review B}\ }\textbf
  {\bibinfo {volume} {100}},\ \bibinfo {pages} {205425} (\bibinfo {year}
  {2019})}\BibitemShut {NoStop}%
\bibitem [{\citenamefont {Newberger}(1982)}]{newberger1982new}%
  \BibitemOpen
  \bibfield  {author} {\bibinfo {author} {\bibfnamefont {B.~S.}\ \bibnamefont
  {Newberger}},\ }\bibfield  {title} {\bibinfo {title} {New sum rule for
  products of bessel functions with application to plasma physics},\
  }\href@noop {} {\bibfield  {journal} {\bibinfo  {journal} {Journal of
  Mathematical Physics}\ }\textbf {\bibinfo {volume} {23}},\ \bibinfo {pages}
  {1278} (\bibinfo {year} {1982})}\BibitemShut {NoStop}%
\bibitem [{\citenamefont {Watson}(1922)}]{watson1922treatise}%
  \BibitemOpen
  \bibfield  {author} {\bibinfo {author} {\bibfnamefont {G.~N.}\ \bibnamefont
  {Watson}},\ }\href@noop {} {\emph {\bibinfo {title} {A treatise on the theory
  of Bessel functions}}},\ Vol.~\bibinfo {volume} {2}\ (\bibinfo  {publisher}
  {The University Press},\ \bibinfo {year} {1922})\BibitemShut {NoStop}%
\bibitem [{\citenamefont {Hall}\ \emph {et~al.}(1965)\citenamefont {Hall},
  \citenamefont {Heckrotte},\ and\ \citenamefont {Kammash}}]{hall1965ion}%
  \BibitemOpen
  \bibfield  {author} {\bibinfo {author} {\bibfnamefont {L.~S.}\ \bibnamefont
  {Hall}}, \bibinfo {author} {\bibfnamefont {W.}~\bibnamefont {Heckrotte}},\
  and\ \bibinfo {author} {\bibfnamefont {T.}~\bibnamefont {Kammash}},\
  }\bibfield  {title} {\bibinfo {title} {Ion cyclotron electrostatic
  instabilities},\ }\href@noop {} {\bibfield  {journal} {\bibinfo  {journal}
  {Physical Review}\ }\textbf {\bibinfo {volume} {139}},\ \bibinfo {pages}
  {A1117} (\bibinfo {year} {1965})}\BibitemShut {NoStop}%
\bibitem [{\citenamefont {Wen}\ and\ \citenamefont {Yu}(2009)}]{wen2009landau}%
  \BibitemOpen
  \bibfield  {author} {\bibinfo {author} {\bibfnamefont {X.}~\bibnamefont
  {Wen}}\ and\ \bibinfo {author} {\bibfnamefont {Y.}~\bibnamefont {Yu}},\
  }\bibfield  {title} {\bibinfo {title} {Landau-zener interference in
  multilevel superconducting flux qubits driven by large-amplitude fields},\
  }\href@noop {} {\bibfield  {journal} {\bibinfo  {journal} {Physical Review
  B}\ }\textbf {\bibinfo {volume} {79}},\ \bibinfo {pages} {094529} (\bibinfo
  {year} {2009})}\BibitemShut {NoStop}%
\bibitem [{\citenamefont {Liul}\ \emph {et~al.}(2023)\citenamefont {Liul},
  \citenamefont {Ryzhov},\ and\ \citenamefont
  {Shevchenko}}]{liul2023interferometry}%
  \BibitemOpen
  \bibfield  {author} {\bibinfo {author} {\bibfnamefont {M.}~\bibnamefont
  {Liul}}, \bibinfo {author} {\bibfnamefont {A.}~\bibnamefont {Ryzhov}},\ and\
  \bibinfo {author} {\bibfnamefont {S.}~\bibnamefont {Shevchenko}},\ }\bibfield
   {title} {\bibinfo {title} {Interferometry of multi-level systems:
  rate-equation approach for a charge qudit},\ }\href@noop {} {\bibfield
  {journal} {\bibinfo  {journal} {The European Physical Journal Special
  Topics}\ ,\ \bibinfo {pages} {1}} (\bibinfo {year} {2023})}\BibitemShut
  {NoStop}%
\bibitem [{\citenamefont {Liul}\ and\ \citenamefont
  {Shevchenko}(2023)}]{liul2023rate}%
  \BibitemOpen
  \bibfield  {author} {\bibinfo {author} {\bibfnamefont {M.}~\bibnamefont
  {Liul}}\ and\ \bibinfo {author} {\bibfnamefont {S.}~\bibnamefont
  {Shevchenko}},\ }\bibfield  {title} {\bibinfo {title} {Rate-equation approach
  for multi-level quantum systems},\ }\href@noop {} {\bibfield  {journal}
  {\bibinfo  {journal} {Low Temperature Physics}\ }\textbf {\bibinfo {volume}
  {49}},\ \bibinfo {pages} {96} (\bibinfo {year} {2023})}\BibitemShut {NoStop}%
\bibitem [{\citenamefont {Be{\'c}a}(1980)}]{beca1980orthogonal}%
  \BibitemOpen
  \bibfield  {author} {\bibinfo {author} {\bibfnamefont {H.~O.}\ \bibnamefont
  {Be{\'c}a}},\ }\bibfield  {title} {\bibinfo {title} {An orthogonal set based
  on bessel functions of the first kind/skup ortogonalnih funkcija zasnovan na
  besselovim funkcijama prve vrste},\ }\href@noop {} {\bibfield  {journal}
  {\bibinfo  {journal} {Publikacije Elektrotehni{\v{c}}kog fakulteta. Serija
  Matematika i fizika}\ ,\ \bibinfo {pages} {85}} (\bibinfo {year}
  {1980})}\BibitemShut {NoStop}%
\bibitem [{\citenamefont {Silveri}\ \emph {et~al.}(2017)\citenamefont
  {Silveri}, \citenamefont {Tuorila}, \citenamefont {Thuneberg},\ and\
  \citenamefont {Paraoanu}}]{silveri2017quantum}%
  \BibitemOpen
  \bibfield  {author} {\bibinfo {author} {\bibfnamefont {M.}~\bibnamefont
  {Silveri}}, \bibinfo {author} {\bibfnamefont {J.}~\bibnamefont {Tuorila}},
  \bibinfo {author} {\bibfnamefont {E.}~\bibnamefont {Thuneberg}},\ and\
  \bibinfo {author} {\bibfnamefont {G.}~\bibnamefont {Paraoanu}},\ }\bibfield
  {title} {\bibinfo {title} {Quantum systems under frequency modulation},\
  }\href@noop {} {\bibfield  {journal} {\bibinfo  {journal} {Reports on
  Progress in Physics}\ }\textbf {\bibinfo {volume} {80}},\ \bibinfo {pages}
  {056002} (\bibinfo {year} {2017})}\BibitemShut {NoStop}%
\bibitem [{\citenamefont {Silveri}\ \emph {et~al.}(2015)\citenamefont
  {Silveri}, \citenamefont {Kumar}, \citenamefont {Tuorila}, \citenamefont
  {Li}, \citenamefont {Veps{\"a}l{\"a}inen}, \citenamefont {Thuneberg},\ and\
  \citenamefont {Paraoanu}}]{silveri2015stuckelberg}%
  \BibitemOpen
  \bibfield  {author} {\bibinfo {author} {\bibfnamefont {M.}~\bibnamefont
  {Silveri}}, \bibinfo {author} {\bibfnamefont {K.}~\bibnamefont {Kumar}},
  \bibinfo {author} {\bibfnamefont {J.}~\bibnamefont {Tuorila}}, \bibinfo
  {author} {\bibfnamefont {J.}~\bibnamefont {Li}}, \bibinfo {author}
  {\bibfnamefont {A.}~\bibnamefont {Veps{\"a}l{\"a}inen}}, \bibinfo {author}
  {\bibfnamefont {E.}~\bibnamefont {Thuneberg}},\ and\ \bibinfo {author}
  {\bibfnamefont {G.}~\bibnamefont {Paraoanu}},\ }\bibfield  {title} {\bibinfo
  {title} {St{\"u}ckelberg interference in a superconducting qubit under
  periodic latching modulation},\ }\href@noop {} {\bibfield  {journal}
  {\bibinfo  {journal} {New Journal of Physics}\ }\textbf {\bibinfo {volume}
  {17}},\ \bibinfo {pages} {043058} (\bibinfo {year} {2015})}\BibitemShut
  {NoStop}%
\bibitem [{\citenamefont {Aspelmeyer}\ \emph {et~al.}(2014)\citenamefont
  {Aspelmeyer}, \citenamefont {Kippenberg},\ and\ \citenamefont
  {Marquardt}}]{aspelmeyer2014cavity}%
  \BibitemOpen
  \bibfield  {author} {\bibinfo {author} {\bibfnamefont {M.}~\bibnamefont
  {Aspelmeyer}}, \bibinfo {author} {\bibfnamefont {T.~J.}\ \bibnamefont
  {Kippenberg}},\ and\ \bibinfo {author} {\bibfnamefont {F.}~\bibnamefont
  {Marquardt}},\ }\bibfield  {title} {\bibinfo {title} {Cavity optomechanics},\
  }\href@noop {} {\bibfield  {journal} {\bibinfo  {journal} {Reviews of Modern
  Physics}\ }\textbf {\bibinfo {volume} {86}},\ \bibinfo {pages} {1391}
  (\bibinfo {year} {2014})}\BibitemShut {NoStop}%
\bibitem [{\citenamefont {Wilson-Rae}\ \emph {et~al.}(2004)\citenamefont
  {Wilson-Rae}, \citenamefont {Zoller},\ and\ \citenamefont
  {Imamoḡlu}}]{wilson2004laser}%
  \BibitemOpen
  \bibfield  {author} {\bibinfo {author} {\bibfnamefont {I.}~\bibnamefont
  {Wilson-Rae}}, \bibinfo {author} {\bibfnamefont {P.}~\bibnamefont {Zoller}},\
  and\ \bibinfo {author} {\bibfnamefont {A.}~\bibnamefont {Imamoḡlu}},\
  }\bibfield  {title} {\bibinfo {title} {Laser cooling of a nanomechanical
  resonator mode to its quantum ground state},\ }\href@noop {} {\bibfield
  {journal} {\bibinfo  {journal} {Physical review letters}\ }\textbf {\bibinfo
  {volume} {92}},\ \bibinfo {pages} {075507} (\bibinfo {year}
  {2004})}\BibitemShut {NoStop}%
\bibitem [{\citenamefont {Ficek}\ \emph {et~al.}(2001)\citenamefont {Ficek},
  \citenamefont {Seke}, \citenamefont {Soldatov},\ and\ \citenamefont
  {Adam}}]{ficek2001fluorescence}%
  \BibitemOpen
  \bibfield  {author} {\bibinfo {author} {\bibfnamefont {Z.}~\bibnamefont
  {Ficek}}, \bibinfo {author} {\bibfnamefont {J.}~\bibnamefont {Seke}},
  \bibinfo {author} {\bibfnamefont {A.}~\bibnamefont {Soldatov}},\ and\
  \bibinfo {author} {\bibfnamefont {G.}~\bibnamefont {Adam}},\ }\bibfield
  {title} {\bibinfo {title} {Fluorescence spectrum of a two-level atom driven
  by a multiple modulated field},\ }\href@noop {} {\bibfield  {journal}
  {\bibinfo  {journal} {Physical Review A}\ }\textbf {\bibinfo {volume} {64}},\
  \bibinfo {pages} {013813} (\bibinfo {year} {2001})}\BibitemShut {NoStop}%
\bibitem [{\citenamefont {Cohen-Tannoudji}(1977)}]{cohen1977frontiers}%
  \BibitemOpen
  \bibfield  {author} {\bibinfo {author} {\bibfnamefont {C.}~\bibnamefont
  {Cohen-Tannoudji}},\ }\bibfield  {title} {\bibinfo {title} {Frontiers in
  laser spectroscopy},\ }in\ \href@noop {} {\emph {\bibinfo {booktitle} {Proc.
  27th Les Houches Summer School}}}\ (\bibinfo {organization} {North-Holland
  Amsterdam},\ \bibinfo {year} {1977})\ p.~\bibinfo {pages} {3}\BibitemShut
  {NoStop}%
\bibitem [{\citenamefont {Thomann}(1980)}]{thomann1980optical}%
  \BibitemOpen
  \bibfield  {author} {\bibinfo {author} {\bibfnamefont {P.}~\bibnamefont
  {Thomann}},\ }\bibfield  {title} {\bibinfo {title} {Optical resonances in a
  strong modulated laser beam},\ }\href@noop {} {\bibfield  {journal} {\bibinfo
   {journal} {Journal of Physics B: Atomic and Molecular Physics}\ }\textbf
  {\bibinfo {volume} {13}},\ \bibinfo {pages} {1111} (\bibinfo {year}
  {1980})}\BibitemShut {NoStop}%
\bibitem [{\citenamefont {Sen}(1952)}]{sen1952solar}%
  \BibitemOpen
  \bibfield  {author} {\bibinfo {author} {\bibfnamefont {H.~K.}\ \bibnamefont
  {Sen}},\ }\bibfield  {title} {\bibinfo {title} {Solar enhanced radiation and
  plasma oscillations},\ }\href@noop {} {\bibfield  {journal} {\bibinfo
  {journal} {Physical Review}\ }\textbf {\bibinfo {volume} {88}},\ \bibinfo
  {pages} {816} (\bibinfo {year} {1952})}\BibitemShut {NoStop}%
\end{thebibliography}%

\end{document}